\documentclass[aps,prb,twocolumn,superscriptaddress,longbibliography]{revtex4-2}

\usepackage[colorlinks=true,citecolor=blue]{hyperref}
\usepackage{graphicx}
\graphicspath{ {./Figures/} }
\usepackage{amsmath}
\usepackage{amssymb}
\usepackage{bbold}
\usepackage{enumerate}
\usepackage[dvipsnames]{xcolor}
\usepackage{physics}

\usepackage{orcidlink}

\begin{document}

\author{Aleksander Sanjuan Ciepielewski\orcidlink{https://orcid.org/0000-0001-5883-4903}}
\affiliation{International Research Centre MagTop, Institute of Physics, Polish Academy of Sciences, Al. Lotnik\'ow 32/46, 02-668 Warsaw, Poland}

\author{Jakub Tworzyd\l{}o\orcidlink{https://orcid.org/0000-0003-3410-5460}}
\affiliation{Faculty of Physics, University of Warsaw, ulica Pasteura 5, 02-093 Warsaw, Poland}

\author{Timo Hyart\orcidlink{https://orcid.org/0000-0003-2587-9755}}
\affiliation{Department of Applied Physics, Aalto University, 00076 Aalto, Espoo, Finland}
\affiliation{Computational Physics Laboratory, Physics Unit, Faculty of Engineering and Natural Sciences, Tampere University, FI-33014 Tampere, Finland}

\author{Alexander Lau\orcidlink{https://orcid.org/0000-0001-6671-8056}}
\affiliation{International Research Centre MagTop, Institute of Physics, Polish Academy of Sciences, Al. Lotnik\'ow 32/46, 02-668 Warsaw, Poland}

\title{Transport effects of twist-angle disorder in mesoscopic twisted bilayer graphene}

\date{\today}


\begin{abstract}

Magic-angle twisted bilayer graphene is a tunable material with remarkably flat energy bands near the Fermi level, leading to fascinating transport properties and correlated states at low temperatures. However, grown pristine samples of this material tend to break up into landscapes of twist-angle domains, strongly influencing the physical properties of each individual sample. This poses a significant problem to the interpretation and comparison between measurements obtained from different samples. In this work, we study numerically the effects of twist-angle disorder on quantum electron transport in mesoscopic samples of magic-angle twisted bilayer graphene. We find a significant property of twist-angle disorder that distinguishes it from onsite-energy disorder: it leads to an asymmetric broadening of the energy-resolved conductance. The magnitude of the twist-angle variation has a strong effect on conductance, while the number of twist-angle domains is of much lesser significance. We further establish a relationship between the asymmetric broadening and the asymmetric density of states of twisted bilayer graphene at angles smaller than the first magic angle. Our results show that the qualitative differences between the types of disorder in the energy-resolved conductance of twisted bilayer graphene samples can be used to characterize them at temperatures above the critical temperatures of the correlated phases, enabling systematic experimental studies of the effects of the different types of disorders also on the other properties such as the competition of the different types of correlated states appearing at lower temperatures.

\end{abstract}

\maketitle

\section{Introduction}

Twisted bilayer graphene (TBG) is a fascinating two-dimentional (2D) material that exhibits  exceptionally flat energy bands with nontrivial topological properties and van Hove singularities near the Fermi level at certain \emph{magic} twist angles~\cite{morell2010, Bistritzer2011a,  Liu2019, Yuan2019, Song2019, Tarnopolsky2019}, leading to an appearance of superconductivity and other correlated phases at low temperatures due to the large density of states~\cite{Kim2017, Ojajarvi2018, Cao2018a, Cao2018b, Liang2018, Wu2018, PeltonenPhysRevB.98.220504,  Lu2019,  Liang2019, Yankowitz2019,  Lu2019, Hazra2019, Hu2019, Fang2020, Julku2020,  Andrei2020} and strongly energy-dependent transport due to the van Hove singularities \cite{Sanjuan2022}. The presence of such rich phenomena in a simple carbon-based 2D structure with a tunable parameter makes TBG one of the best model systems for investigating the exotic flat-band physics.

Experimental and theoretical studies have shown that magic-angle TBG is very sensitive to the twist angle. In particular, the width of the quasi-flat energy bands and their overall topology changes abruptly with variations of even a tenth of a degree of twisting. Moreover, the TBG samples typically crystallize into unique landscapes of domains with slightly varying twist angles~\cite{Uri2020, deJong2022, Hu_2024}, and with the current fabrication techniques the twist-angle variation can be reduced down to a range of $\delta \theta \approx 0.1^\circ$~\cite{Uri2020}.  Therefore, it is important to understand the effect of twist-angle disorder on the physical properties of TBG. Previous theoretical works on twist-angle disorder in TBG have used a variety of approaches, such as a real-space domain models~\cite{Wilson2020, Cruz2021}, non-uniform moir{\'e} patterns with a minimal continuous model~\cite{Naoto2022}, transmission calculations though one-dimensional variation of twist angle in minimal continuous models~\cite{Padhi20, Joy2020, Cruz2021}, and a Landau-Ginzburg theory to study the interplay between electron-electron interactions and disorder~\cite{Thomson2021}. In this work we introduce a novel method for the study of twist-angle disorder in TBG, in which we induce disorder via the interlayer hopping amplitude instead of directly through the twist angle. This approach is possible due to the equivalence of small deviations in twist-angle and  interlayer hopping amplitude discussed in our previous work~\cite{Sanjuan2022}.

Most of the research on TBG has so far focused on observables in the thermodynamic limit and on transport in macroscopic samples in the semiclassical regime ~\cite{WudasSarma19, HwangdasSarma}. Previous quantum transport studies of TBG have addressed specific questions, such as the angle-dependent minimal conductivity and effects of onsite disorder ~\cite{Andelkovic2018, Hou2024}, transport across twist-angle domains ~\cite{Padhi20}, and emergent magnetic textures in driven TBG~\cite{Bahamon2020}. Studies on quantum transport have often used crossed graphene nanoribbons ~\cite{Zhou2010, Brandimarte2017, Sanz2020}, where the scattering region is smaller than the magic-angle moir{\'e} unit cell. In our previous work we showed that quantum  transport in mesoscopic TBG samples allow to probe the electronic properties in an energy-resolved fashion, revealing detailed effects arising from the quasi-flat band topology and van Hove singularities~\cite{Sanjuan2022}. Such kind of quantum transport calculations can retain information about coherent quantum effects  over the length scale of many magic-angle moir{\'e} unit cells. 

In this paper we study the effects of twist-angle disorder on the two-terminal conductance in mesoscopic TBG samples containing approximately one million sites, near the first magic angle $\theta_m \approx 1.05^\circ$. We first show that small deviations of twist angle and interlayer hopping amplitude induce the same changes to the topology of the moir{\'e} bands and to the electron density of states (DOS), and the interval of twist angles in which this equivalence takes place is larger than the variation of the twist angle in the state-of-the-art experimental samples.  
Then we demonstrate that the effects of twist-angle disorder are qualitatively different from onsite-energy disorder by comparing the energy-resolved conductances. Finally, we establish a connection between the differences observed in the conductances and  DOS in the presence of these two types of disorder. Our findings show that conductance measurements can be used to quantify the type and strength of disorder present in TBG samples at temperatures higher than the critical temperatures of the correlated phases, enabling comparisons between measurements obtained from different samples and systematic experimental studies of the dependencies of the various physical properties on the disorder. 

The structure of the paper is as follows. In Sec.~\ref{sec:model-and-setup} we introduce the two-terminal geometry for quantum transport calculations and the tight-binding model. In Sec.~\ref{sec:equivalence} we show the local equivalence between the variation of twist angle and interlayer hopping amplitude in magic-angle TBG, and describe how we use this equivalence to induce disorder in the samples. In Sec.~\ref{sec:conductance-disorder} we show the effects of twist-angle disorder and onsite-energy disorder on the energy-resolved conductance as a function of disorder strength and number of disorder domains. In Sec.~\ref{sec:dos_disorder} we present the effects of twist-angle and onsite-energy disorder on the DOS, and show how they relate to the conductance results. Finally, we summarize our results in Sec.~\ref{sec:conclusion}.

\section{Model and setup}
\label{sec:model-and-setup}

\begin{figure}
\includegraphics[width=0.485\textwidth]{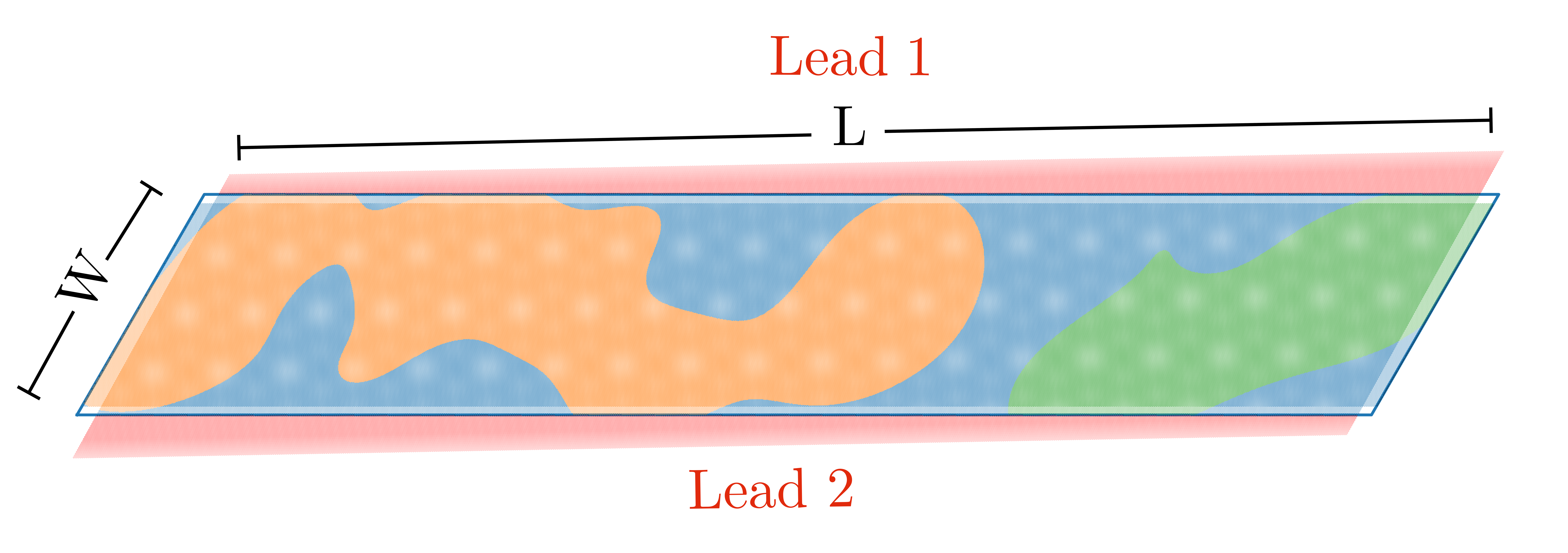}
\vspace{-0.5cm}
\caption{Twisted bilayer graphene setup: two crossed graphene nanoribbons with leads (red) form a bilayer region. The top layer is twisted relative to the bottom layer around the center of the overlap region by an angle $\theta \approx 1.05^\circ$. Each color in the sample represents a twist-angle domain. We show a system of width $W \approx 50 \mathrm{nm}$ and length $L \approx 250 \mathrm{nm}$, which is the actual size of the systems considered in this work.}
\label{fig:setup}
\end{figure}

To study quantum transport through a TBG region formed by two crossed graphene ribbons, as illustrated in Fig.~\ref{fig:setup}, we utilize similar approach as in our previous work~\cite{Sanjuan2022}. The ribbons are placed such that for $\theta=0$ the two layers are stacked in a AA fashion, and then they are 
rotated relative to each other by an angle around the center of the overlap region. The continuation of the top ribbon outside the bilayer region defines one semi-infinite, monolayer graphene lead, which we use for our transport calculations (see red section in Fig.~\ref{fig:setup}). In contrast to Ref.~\cite{Olyaei2020}, we make the leads metallic by setting the chemical potential of the leads far away from the Dirac-point energy, and deep into the bulk of the spectrum, to contain as many modes as possible. As shown in our previous work \cite{Sanjuan2022}, this setup realizes a short and wide contact capable of probing the properties of the system in an energy resolved fashion, thanks to the large number (thousands) of modes in the leads. 

The low-energy properties of TBG have been studied using continuum models~\cite{dosSantos2007,Mele2010,Bistritzer2011a}, ab-initio calculations~\cite{Sanz2020, Zhou2010, Brandimarte2017}, and tight-binding models~\cite{Cao2021, morell2010, Lin2018, Moon2012, Po2019}. Continuum models are long-wavelength, low-energy theories and, therefore, cannot fully capture the details at lengths and energy scales relevant to transport in mesoscopic systems. The ab-initio calculations are computationally expensive to model samples sufficiently large to overcome finite-size effects at small twist angles, and thereby the effects attributable to the electronic properties of the bulk are necessarily obscured. Therefore, in this work we adopt the tight-binding model approach, which is  able to accurately capture the low-energy electronic properties of TBG over a wide range of twist angles in the vicinity of the first magic angle~\cite{Lin2018, Moon2012} and allows to efficiently study systems with close to $10^6$ lattice sites. In our conductance calculations we used samples of $4\times 20$ magic-angle moir\'{e} unit cells, which is equivalent to $50\,\mathrm{nm} \times 250\,\mathrm{nm}$.

The  Hamiltonian for TBG is $H = H_1 + H_2 + H_{12}$, where $H_{1,2}$ are the nearest-neighbor tight-binding Hamiltonians of the individual layers~\cite{CastroNeto2009}
\begin{eqnarray}
    H_m &=& -t\sum_{\langle i,j\rangle,\sigma} c_{im\sigma}^\dagger c_{jm\sigma} - \mu \sum_{i,\sigma} c_{im\sigma}^\dagger c_{im\sigma},
\end{eqnarray}
and $H_{12}$ contains interlayer-coupling terms~\cite{Moon2012, Olyaei2020, Lin2018}
\begin{equation}
    H_{12} = -\sum_{\langle i,j\rangle,\sigma} t'(r_{ij})\,c_{i,2,\sigma}^\dagger c_{j,1,\sigma} + \mathrm{H.c.},
\end{equation}
where $r_{ij}=|\mathbf{r}_i-\mathbf{r}_j|$ is the in-plane distance between two lattice sites in different layers at positions $\mathbf{r}_i$ and $\mathbf{r}_j$, respectively, and $t'(r)$ is the isotropic interlayer hopping integral given by
\begin{equation}
    t'(r) = V_{pp\sigma}^0\, e^{-\left(\sqrt{r^2+d_0^2}-d_0\right)/\lambda} \frac{d_0^2}{r^2+d_0^2}.
    \label{eq:interlayer_coupling}
\end{equation}
Here $c_{im\sigma}^\dagger$ ($c_{im\sigma}$) creates (annihilates) a $p_z$ electron with spin $\sigma=\uparrow,\downarrow$ at lattice site $\mathbf{r}_i$ of the $m$-th layer ($m=1,2$), $t=3.09\,\mathrm{eV}$ is the nearest-neighbor hopping amplitude~\cite{Lin2018}, $\mu$ is the chemical potential,  $V^0_{pp\sigma} = 0.39\,\mathrm{eV}$ is the nearest-neighbor interlayer coupling, $d_0=3.35\,\textrm{\AA}$ is the distance between the graphene layers, and  $\lambda$ is the decay parameter.
In accordance with Ref.~\onlinecite{Lin2018}, we use $\lambda=0.27\,\textrm{\AA}$, which reproduces the band structures of AA- and AB-stacked BLG. Making use of the rapidly decaying nature of the hopping integral $t(r)$ in Eq.~\eqref{eq:interlayer_coupling}, we further neglect interlayer terms with $r > 5\,\textrm{\AA}$, which is sufficient to accurately capture the moir\'{e} bands of TBG near the first-magic angle. We set $\mu=2\,\mathrm{eV}$ in the leads, such that the leads are metallic, and $\mu=0$ throughout the scattering region. In our calculations we use the quantum transport Python package kwant~\cite{Groth2014}. 

We further aim to draw a connection between transport signatures and spectral features of the bulk. For a countable set of commensurate twist angles, the emerging moir{\'e} pattern is periodic, and forms a hexagonal superlattice, so that we can impose periodic boundary conditions on the moir\'{e} unit cell and calculate the moir{\'e} band structure, from which we can extract the bulk DOS. The set of commensurate twist angles has an accumulation point at zero, thus at small angles $\theta \approx 1^\circ$ we find a commensurate twist angle in the vicinity ($\delta \theta \lesssim 0.1^\circ$) of any angle. In conductance calculations we use a commensurate twist angle $\theta_m = \mathrm{arccos}(2976.5 / 2977)^\circ \approx 1.05^\circ$, which is located near the experimentally discovered magic-angle $\theta \approx 1.1^\circ$~\cite{Cao2018a}. At $\theta_m$ the moir\'{e} lattice constant is $ a\approx  15\,\textrm{nm}$, the moir\'{e} bands are remarkably flat (bandwidth $\Delta E \approx 3 \, \mathrm{meV}$), and we obtain the strongest conductance signal~\cite{Sanjuan2022}, but we note that the smallest bandwidth of the moir\'{e} bands in our model occurs at the commensurate twist angle $\theta = \mathrm{arccos}(3168.5 / 3169)^\circ \approx 1.02^\circ$, which is also used in our DOS calculations, similarly as the commensurate twist angle $\theta = \mathrm{arccos}(2790.5 / 2791)^\circ\approx1.08^\circ$ (see below).
We emphasize that the commensurate angles are only used to enable efficient calculation of the DOS. The conductance does not depend on whether the system is commensurate or incommensurate. Throughout this work, we align $E=0$ with the energy of the Dirac points at the $\mathrm{K}$ and $\mathrm{K}'$ points of the moir{\'e} Brillouin zone (BZ).

 We restrict our study to a non-interacting description of the system, which is a good approximation as long as the Fermi level is tuned away from the quasi-flat bands or temperature is above the critical temperature of the correlated phases. We note that even if the Fermi level is tuned away from the quasi-flat bands, the transport can still be studied in an energy-resolved manner across the flat band energies by tuning the voltage bias. In this case, only nonequilibrium quasiparticles are occupying the flat-band states so that interaction effects are not expected to be as important as in the case of equilibrium flat-band systems. Alternatively, it is also possible to screen the interactions so that a non-interacting description becomes more accurate \cite{Veyrat2020, Stepanov2020}.  
From the practical perspective, it is easiest to avoid the interaction-induced reconstruction of the energy bands, which obscures the effects of van Hove singularities and the disorder on transport, by increasing the temperature above the critical temperature of the correlated phases. Therefore, in the following we pay special attention also to the temperature dependence of the conductance.

\begin{figure}
\includegraphics[width=0.5\textwidth]{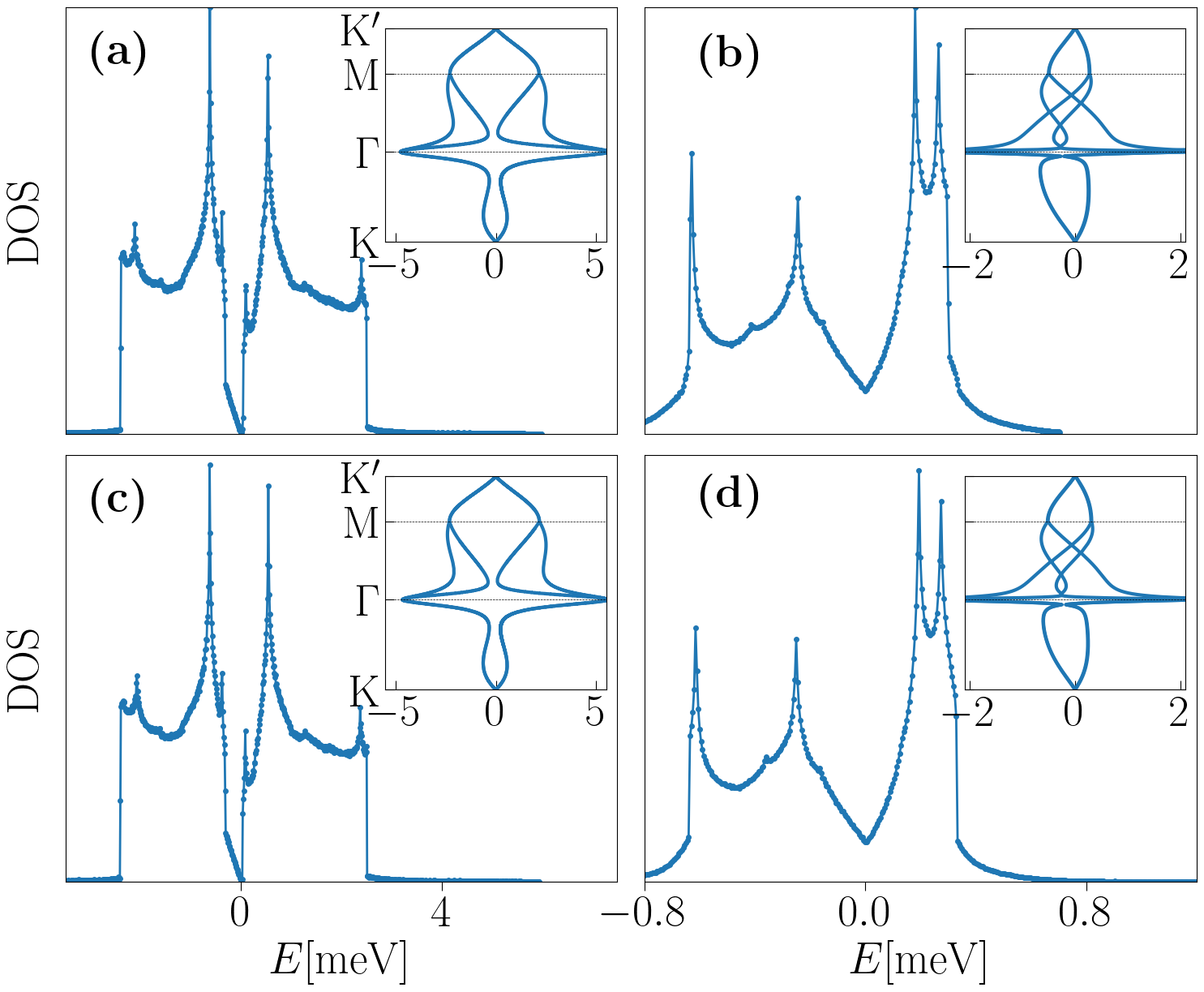}
\caption{(a),(b) DOS and spectrum through high-symmetry lines for $\theta \approx 1.08^\circ$ and $\theta \approx 1.02^\circ$, when the interlayer hopping amplitude is fixed to $V^0_{pp\sigma}=390 \mathrm{meV}$. (c),(d) Same for interlayer hopping amplitudes  $V^0_{pp\sigma}=377 \mathrm{meV}$ and $V^0_{pp\sigma}=403 \mathrm{meV}$, when the twist angle is fixed to  $\theta_m \approx 1.05^\circ$.}
\label{fig:equivalence}
\end{figure}

\section{Effective model for twist-angle disorder}
\label{sec:equivalence}

We begin by discussing the local equivalence between twist angle and interlayer hopping amplitude in TBG. Panels (a) and (b) in Fig.~\ref{fig:equivalence} show the DOS and the moir\'{e} bands through high-symmetry lines of TBG at the two commensurate angles closest to the first magic angle, at a fixed interlayer hopping amplitude. Panels (c) and (d) show the same at the first magic angle, with slightly changed interlayer hopping amplitudes. The width of the moir\'{e} bands, the position of van Hove singularities (shown in Fig.~\ref{fig:equivalence} as peaks in the DOS) and the overall topology of the bands is near identical. Thus, we can evolve the moir\'{e} bands starting from the first magic angle to its two nearest commensurate twist angles, one below and one above it, either using twist angle or the interlayer hopping amplitude as a parameter. Roughly speaking, the twist angle in TBG tunes the coupling between the layers. Indeed, for large angles $\theta \gtrapprox 3^\circ$ the two layers are effectively decoupled, and consequently the system has a conductance equal to twice that of monolayer graphene. On the other hand, at small angles the graphene sheets show enhanced conductance, and near $\theta \approx 1^\circ$ the graphene sheets are strongly coupled~\cite{Andelkovic2018, Sanjuan2022}. Thus, it is not surprising that the twist angle has a similar effect as the interlayer hopping amplitude, which by definition determines the strength of the interlayer coupling. However, it is remarkable that also the changes in the topology of the moir\'{e} bands are effectively identical in the vicinity of the magic angle.

Using this approach we can establish an equivalence between small twist-angle deviations $\delta \theta$ and interlayer hopping amplitude deviations $\delta V_{pp\sigma}^0$. In our model $\delta \theta=0.1^{\circ}$ corresponds to $\delta V_{pp\sigma}^0 = 42.5 \mathrm{meV}$. Furthermore, we find that around $\theta_m$ this equivalence holds for deviations in twist angle of more than $\delta \theta = 0.1^{\circ}$, which is the reported range of twist-angle disorder in pristine samples of TBG using state-of-the-art experimental crystal growth techniques~\cite{Uri2020}. We note that the center of the moir\'{e} bands in energy is slightly different when varying $V_{pp\sigma}^0$ instead of $\theta$. However, these shifts are so small that they do not play a significant role in our analysis. Thus, for spectrum calculations we fix the center of the moir\'{e} bands at $E = 0$. 

We aim to study the effects of disorder in conductance, both with regard to disorder strength and number of twist-angle domains throughout the sample. 
For this purpose we generated 
smooth bubble-like domains (see Fig.~\ref{fig:setup}) by utilizing the contour lines  $g(\vec{x}) = C$ of a scalar function $g(\vec{x})$
consisting of a sum of 2D Gaussian functions, where the centers of the Gaussians are sampled from uniform random distribution (see Appendix~\ref{app:generation_samples}). For one set of parameters (twist angle, interlayer hopping amplitude, and number of domains) we generate twenty samples with randomized twist-angle domains. Then we simulate twist-angle disorder in our samples by utilizing the equivalence between twist angle and interlayer hopping amplitude discussed above. Thus, we fix  the twist angle to $\theta_m \approx1.05^\circ$, and induce disorder by deviating the interlayer hopping amplitude in the twist-angle domains as $V_{pp\sigma}^0 \rightarrow V_{pp\sigma}^0 + \delta V$, where $\delta V$ are sampled from the uniform random distribution $U(-\delta V_{pp\sigma}, \delta V_{pp\sigma})$. Thus, each sample in the ensemble has a distinctive landscape of domains and each domain has a slightly varying interlayer hopping amplitude, as exemplified in Fig. \ref{fig:setup}. When calculating the interlayer hopping between sites belonging to different domains, we use the average of the interlayer hopping amplitudes in each domain. 

\section{Effects of disorder and temperature on conductance}
\label{sec:conductance-disorder}

For a two-terminal setup, the differential conductance $G$ is defined through the relation $G =dI/dV$, where $I$ is the current and $V$ is the applied voltage. We are interested in the conductance $G(E)$ as a function of energy $E=E_f+eV$ (controlled by  $V$) throughout the energy window of the quasi-flat bands at the first magic angle, which in our model has width $\Delta E \approx 3 \mathrm{meV}$. In the Landauer-Büttiker formalism, the conductance $G(E)$ is related to the total transmission $T(E)$ between the terminals  (sum over the transmissions probabilities between all incoming and outgoing modes) through the relation
\begin{equation}
    G(E) = 2\frac{e^2}{h} \int d\tilde{E} \bigg(- \frac{\partial f(\tilde{E} - E)}{\partial \tilde{E}} \bigg)T(\tilde{E}), 
\label{eq:coductance_temperature}
\end{equation}
where $f(E) = 1 / (e^{\beta E} + 1)$ is the Fermi distribution, $\beta=1/(k_B T)$ and the factor $2$ originates from the spin degeneracy. At zero temperature, $-\partial f(E) / \partial E  = \delta(E)$, and thus $G(E) = (2e^2 / h) T(E)$. At nonzero temperature the derivative of the Fermi function in Eq.~(\ref{eq:coductance_temperature}) has a sharp peak centered at $E$ and width proportional to $k_B T$. Thus, the conductance $G(E)$ is determined by the weighted average of the total transmission $T(E)$ over an energy window proportional to the thermal energy. 

\subsection{Effects of twist-angle disorder}

\begin{figure}
\includegraphics[width=0.48\textwidth]{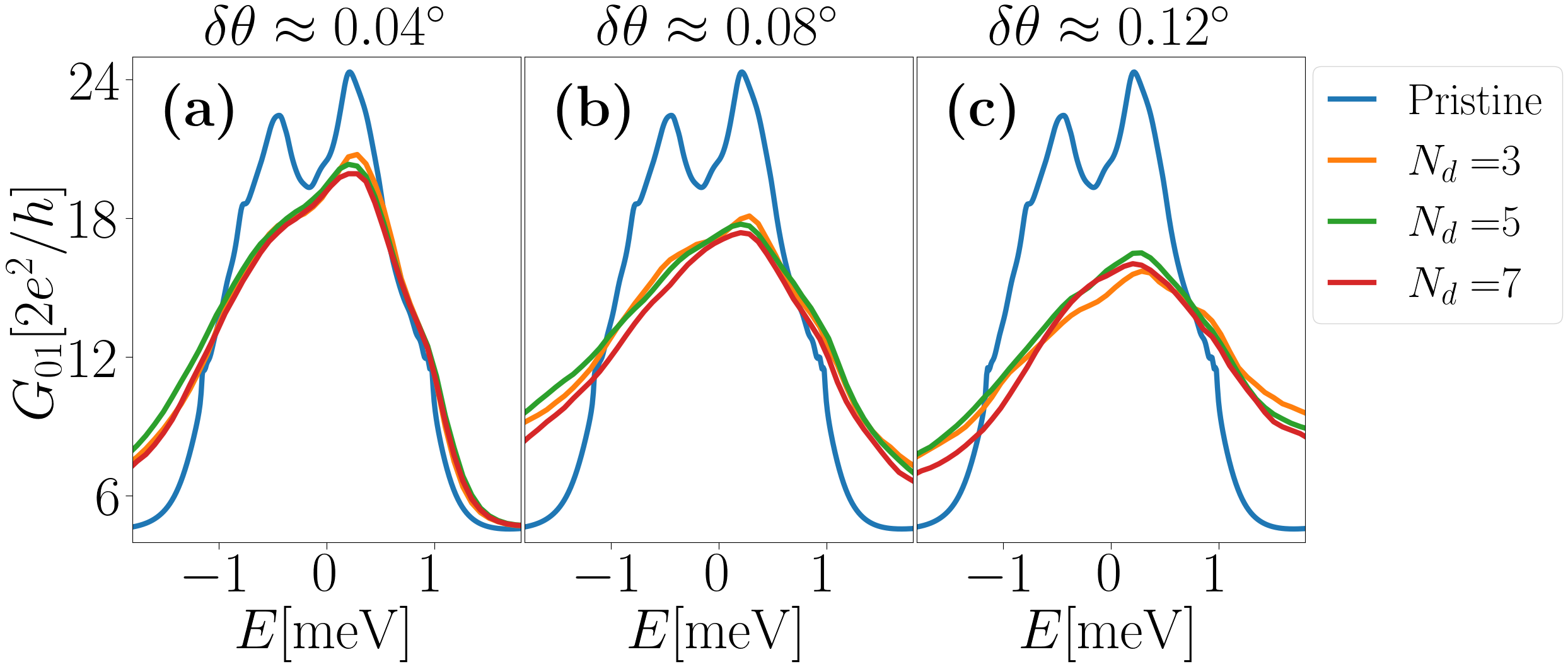}
\caption{Energy-resolved disorder-averaged conductance $G(E)$ for  $N_d=:3, 5$, and $7$ twist-angle domains. The strengths of the twist-angle disorder are: (a) $\delta \theta \approx 0.04^\circ$ (corresponding to $\delta V^0_{pp\sigma} = 17\mathrm{meV}$), (b) $\delta \theta \approx 0.08^\circ$ ($\delta V^0_{pp\sigma} = 34 \mathrm{meV}$) and (c) $\delta \theta \approx 0.12^\circ$ ($\delta V^0_{pp\sigma} = 51 \mathrm{meV}$). In the disorder average we use an ensemble of 20 TBG samples (see Appendix \ref{app:generation_samples}).  
The conductance of the pristine TBG sample is shown for comparison in all figures (blue line).}
\label{fig:coductance_twist_angle_disorder}
\end{figure}

The disorder-averaged conductance $G(E)$ for a range of values of disorder strength $\delta \theta$ (modelled via $\delta V_{pp\sigma}$ as discussed above) and number of domains $N_d$ is shown in Fig.~\ref{fig:coductance_twist_angle_disorder}.  
Already very small disorder strengths [(a) $\delta \theta \approx 0.04^\circ$, (b) $\delta \theta \approx 0.08^\circ$ and (c) $\delta \theta \approx 0.12^\circ$] lead to strong suppression of the conductance  while the number of domains influences the results only weakly. Interestingly, the suppression is asymmetric in energy: The left peak in the conductance of the pristine sample in Fig.~\ref{fig:coductance_twist_angle_disorder}, which corresponds to a van Hove singularity in the hole bands, is quickly washed out, while the right peak coming from the electron bands is more robust. As we will show  in Sec.~\ref{fig:dos_disorder}, this is due to the asymmetry of the DOS of TBG below the first magic twist angle.

\subsection{Effect of onsite-energy disorder}

\begin{figure}
\includegraphics[width=0.45\textwidth]{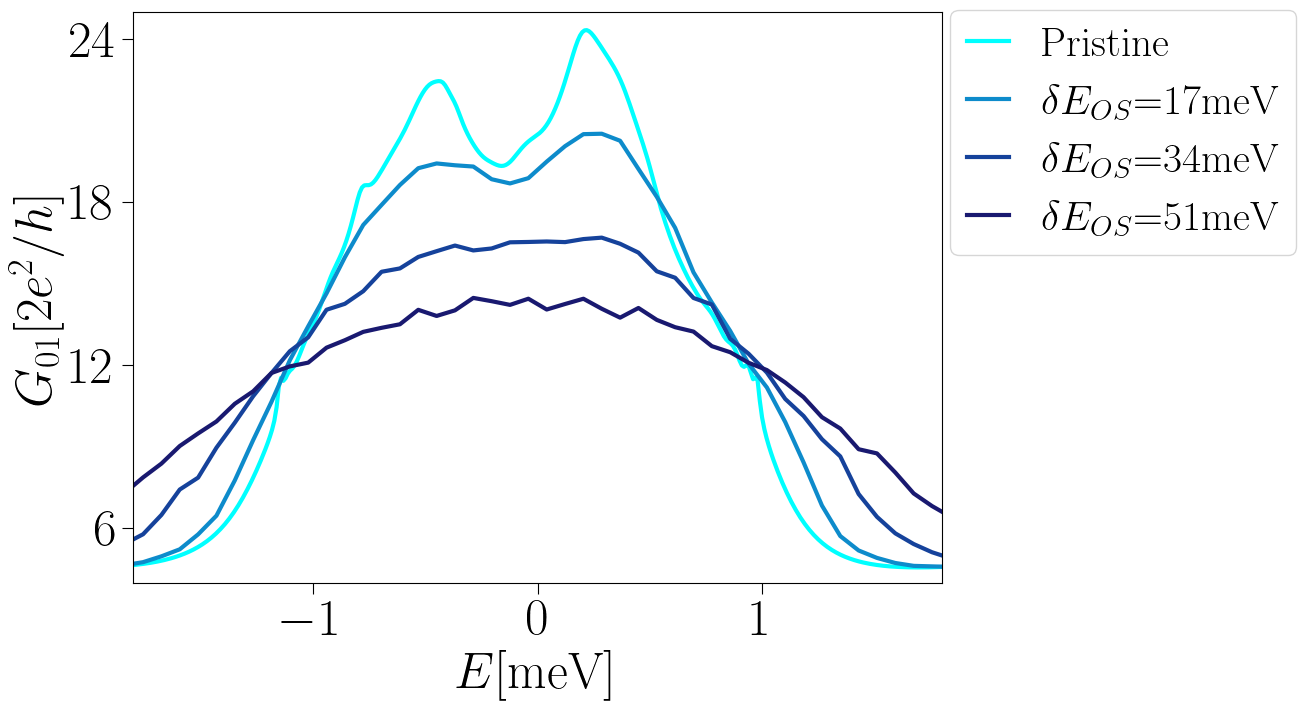}
\caption{Energy-resolved disorder-averaged conductance $G(E)$ for different values of disorder strength $\delta E_{OS}$.  In the disorder average we use an ensemble of 20 TBG samples.  The conductance of the pristine TBG sample is shown for comparison.}
\label{fig:conductance_onsite_energy_disorder}
\end{figure}

In order to highlight that the asymmetric effect of the twist-angle disorder is a particularity of the type of disorder, we calculated the conductance $G(E)$ for magic-angle TBG also in the presence of onsite-energy disorder, see  Fig.~\ref{fig:conductance_onsite_energy_disorder}. In this case the broadening and suppression of the conductance peaks is clearly symmetric, and for sufficiently strong disorder the two peaks merge so that maximum conductance appears in the middle of them. 
\subsection{Effect of temperature}

\begin{figure}
\includegraphics[width=0.5 \textwidth]{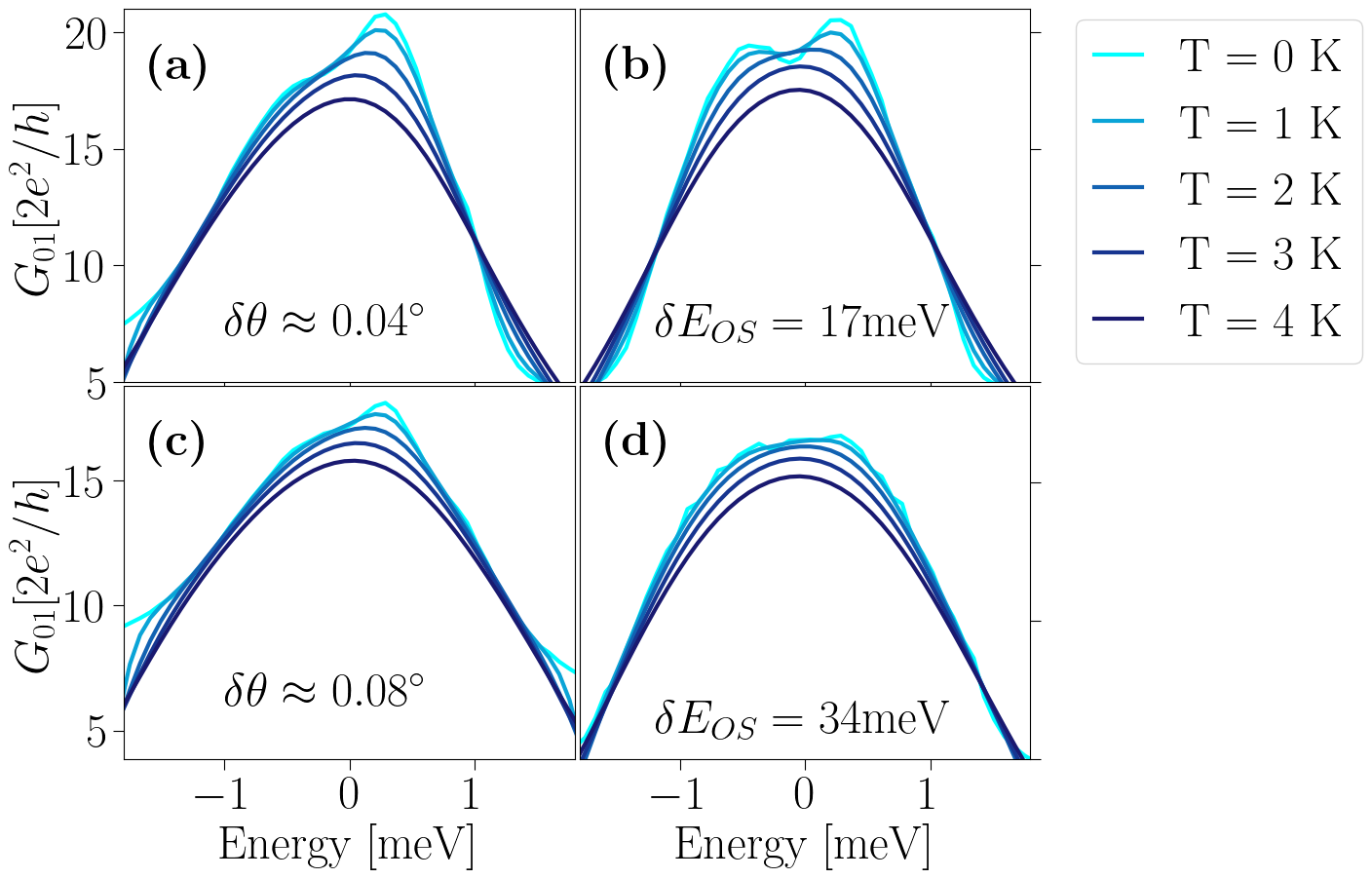}
\caption{Disorder-averaged conductance $G(E)$ of magic-angle TBG samples at different  temperatures for twist-angle disorder strengths (a) $\delta \theta \approx 0.04^\circ$ ($\delta V^0_{pp\sigma} = 17\mathrm{meV}$) and (c) $\delta \theta \approx 0.08^\circ$ ($\delta V^0_{pp\sigma} = 34 \mathrm{meV}$),
and same for onsite-energy disorder strengths (b) $\delta E_{OS}= 17 \mathrm{meV}$ and (d) $\delta E_{OS} = 34 \mathrm{meV}$.}
\label{fig:conductance_temperature}
\end{figure}

Figure \ref{fig:conductance_temperature} shows the temperature-dependence of the disorder-averaged conductance $G(E)$ calculated using Eq.~(\ref{eq:coductance_temperature}). The asymmetric suppression of the conductance in the presence of twist-angle disorder is visible at temperatures below 3 K [panels (a), (c)], so that these $G(E)$ curves can be clearly distinguished from the $G(E)$ curves in the presence of  onsite-energy disorder [panels (b), (d)].  The typical critical temperatures of the correlated phases in magic-angle TBG are $T_c \approx 1$-$2$ K, which means that the type and the strength of the disorder present in the samples can be characterized at $T>T_c$.

\section{Effects of disorder on the density of states}
\label{sec:dos_disorder}

The conductance peaks in magic-angle TBG are related to van Hove singularities in the DOS~\cite{Sanjuan2022}. Thus, we investigated if the asymmetry in the suppression of the peaks shown in Fig.~\ref{fig:coductance_twist_angle_disorder} is connected to asymmetries in the DOS near the first magic angle. Indeed, Fig~\ref{fig:equivalence} shows that for twist angles below $\theta_m$ (panels (b) and (d)) the DOS of the hole bands is spread wider in energy than the DOS of the electron bands. 

\begin{figure}
\includegraphics[width=0.5 \textwidth]{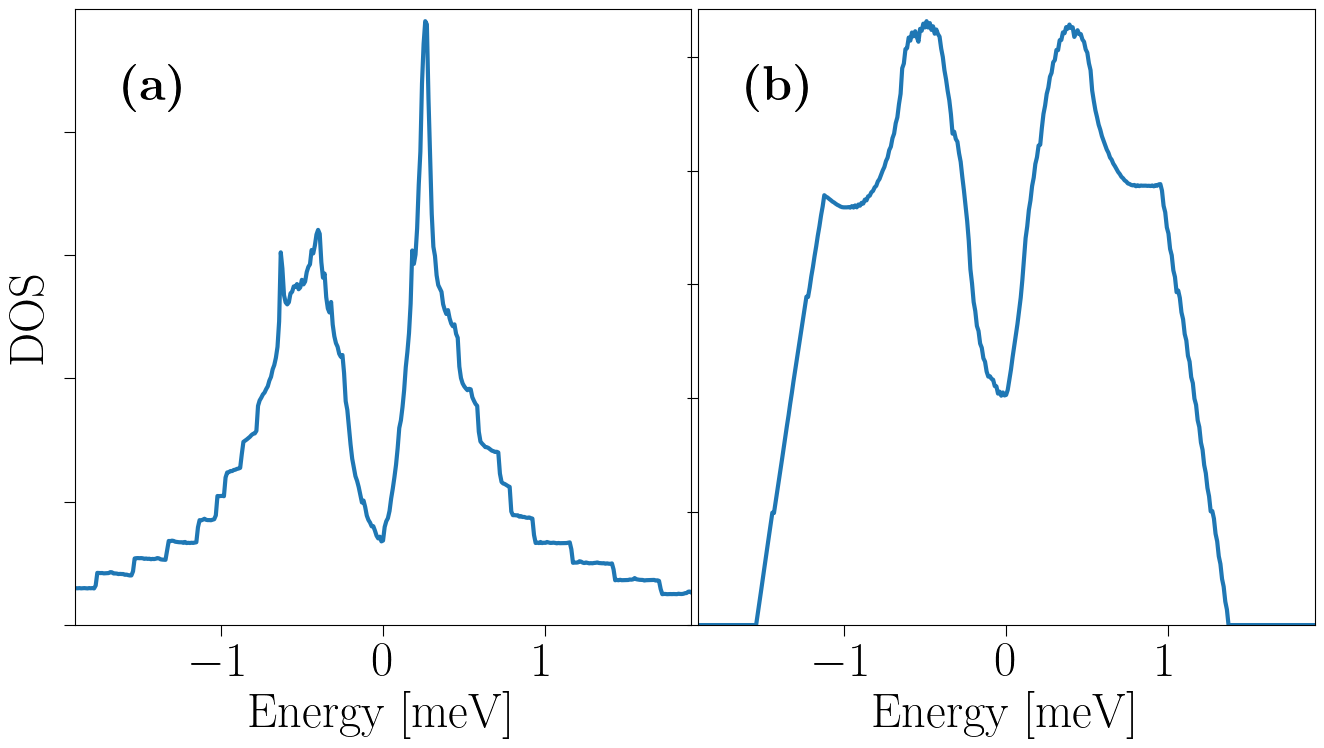}
\caption{DOS of magic-angle TBG samples averaged over (a) an ensemble of samples with interlayer couplings uniformly distributed within the range $V^0_{pp\sigma} \in [377, 403] \ \mathrm{meV}$ and (b) an ensemble of samples with Fermi levels uniformly distributed within the range $E_f \in [-200, 200] \ \mathrm{meV}$.}
\label{fig:dos_disorder}
\end{figure}

Fig~\ref{fig:dos_disorder}(a) shows the DOS averaged over an ensemble of twenty samples with twist angles in the vicinity of the magic angle. The averaged DOS is spread wider and has a smaller peak in the hole bands than in the electron bands. For comparison, if we average the DOS over an ensemble of samples with different Fermi levels  [see Fig.~\ref{fig:dos_disorder} (b)], we find a symmetric averaged DOS. This clearly indicates that the asymmetric suppression of the conductance in magic-angle TBG is a particularity of the twist-angle disorder.

\section{Conclusion}
\label{sec:conclusion}

Twist-angle disorder is thought to be one of the main reasons behind the variations between the properties of the different magic-angle TBG samples. In this paper, we have shown that  the twist-angle disorder leads to significant suppression of the conductance. The conductance depends strongly on the strength of the twist-angle variations but the size of twist-angle domains influences the conductance only weakly. Remarkably, we found that the effect twist-angle disorder is qualitatively different from the effect of onsite-energy disorder in magic-angle TBG. Namely, the twist-angle disorder leads to the asymmetric suppression of the conductance peaks originating from the van Hove singularities in the electron and hole bands, whereas the onsite-energy disorder suppresses these conductance peaks symmetrically.  This difference is a useful tool for the characterization of the type and strength of the disorder present in each magic-angle TBG sample because the effect is observable at temperatures above the critical temperatures of the correlated phases. We further found that the asymmetric suppression of the conductance is a consequence of the asymmetry of the DOS at angles below the magic angle $\theta_m \approx 1.05^\circ$. This is another important finding because it indicates that for angles significantly above $\theta_m$ the twist-angle disorder does not lead to this kind of  asymmetric effects.

\begin{acknowledgments}
A.~S.~acknowledges the initial insight to study twist-angle disorder in the manner described in this work proposed by Jose Lado. 
The research was partially supported by the Foundation for Polish Science through the IRA Programme co-financed by EU within SG OP.  T.~H.~acknowledges the computational resources provided by the Aalto Science-IT project and the financial support from the Academy of Finland Project No.~331094. A.~L.~acknowledges support from a Marie Sk{\l}odowska-Curie Individual Fellowship under grant MagTopCSL (ID 101029345). J.~T.~received funding from the National Science Centre, Poland, within the QuantERA II Programme that has received funding from the European Union’s Horizon 2020 research and innovation programme under Grant Agreement No 101017733, Project Registration Number: 2021/03/Y/ST3/00191, acronym TOBITS. We acknowledge the access to the computing facilities of the Interdisciplinary Center of Modeling at the University of Warsaw, Grant No. G90-1376.

\end{acknowledgments}

\textit{Data availability:} The data shown in the figures is available at Ref.~\onlinecite{zenodo}.

\bibliography{bibliography}

\begin{thebibliography}{51}%
\makeatletter
\providecommand \@ifxundefined [1]{%
 \@ifx{#1\undefined}
}%
\providecommand \@ifnum [1]{%
 \ifnum #1\expandafter \@firstoftwo
 \else \expandafter \@secondoftwo
 \fi
}%
\providecommand \@ifx [1]{%
 \ifx #1\expandafter \@firstoftwo
 \else \expandafter \@secondoftwo
 \fi
}%
\providecommand \natexlab [1]{#1}%
\providecommand \enquote  [1]{``#1''}%
\providecommand \bibnamefont  [1]{#1}%
\providecommand \bibfnamefont [1]{#1}%
\providecommand \citenamefont [1]{#1}%
\providecommand \href@noop [0]{\@secondoftwo}%
\providecommand \href [0]{\begingroup \@sanitize@url \@href}%
\providecommand \@href[1]{\@@startlink{#1}\@@href}%
\providecommand \@@href[1]{\endgroup#1\@@endlink}%
\providecommand \@sanitize@url [0]{\catcode `\\12\catcode `\$12\catcode
  `\&12\catcode `\#12\catcode `\^12\catcode `\_12\catcode `\%12\relax}%
\providecommand \@@startlink[1]{}%
\providecommand \@@endlink[0]{}%
\providecommand \url  [0]{\begingroup\@sanitize@url \@url }%
\providecommand \@url [1]{\endgroup\@href {#1}{\urlprefix }}%
\providecommand \urlprefix  [0]{URL }%
\providecommand \Eprint [0]{\href }%
\providecommand \doibase [0]{https://doi.org/}%
\providecommand \selectlanguage [0]{\@gobble}%
\providecommand \bibinfo  [0]{\@secondoftwo}%
\providecommand \bibfield  [0]{\@secondoftwo}%
\providecommand \translation [1]{[#1]}%
\providecommand \BibitemOpen [0]{}%
\providecommand \bibitemStop [0]{}%
\providecommand \bibitemNoStop [0]{.\EOS\space}%
\providecommand \EOS [0]{\spacefactor3000\relax}%
\providecommand \BibitemShut  [1]{\csname bibitem#1\endcsname}%
\let\auto@bib@innerbib\@empty
\bibitem [{\citenamefont {Su\'arez~Morell}\ \emph {et~al.}(2010)\citenamefont
  {Su\'arez~Morell}, \citenamefont {Correa}, \citenamefont {Vargas},
  \citenamefont {Pacheco},\ and\ \citenamefont {Barticevic}}]{morell2010}%
  \BibitemOpen
  \bibfield  {author} {\bibinfo {author} {\bibfnamefont {E.}~\bibnamefont
  {Su\'arez~Morell}}, \bibinfo {author} {\bibfnamefont {J.~D.}\ \bibnamefont
  {Correa}}, \bibinfo {author} {\bibfnamefont {P.}~\bibnamefont {Vargas}},
  \bibinfo {author} {\bibfnamefont {M.}~\bibnamefont {Pacheco}},\ and\ \bibinfo
  {author} {\bibfnamefont {Z.}~\bibnamefont {Barticevic}},\ }\bibfield  {title}
  {\bibinfo {title} {{Flat bands in slightly twisted bilayer graphene:
  Tight-binding calculations}},\ }\href
  {https://doi.org/10.1103/PhysRevB.82.121407} {\bibfield  {journal} {\bibinfo
  {journal} {Phys. Rev. B}\ }\textbf {\bibinfo {volume} {82}},\ \bibinfo
  {pages} {121407} (\bibinfo {year} {2010})}\BibitemShut {NoStop}%
\bibitem [{\citenamefont {Bistritzer}\ and\ \citenamefont
  {MacDonald}(2011)}]{Bistritzer2011a}%
  \BibitemOpen
  \bibfield  {author} {\bibinfo {author} {\bibfnamefont {R.}~\bibnamefont
  {Bistritzer}}\ and\ \bibinfo {author} {\bibfnamefont {A.~H.}\ \bibnamefont
  {MacDonald}},\ }\bibfield  {title} {\bibinfo {title} {{Moire bands in twisted
  double-layer graphene}},\ }\href {https://doi.org/10.1073/pnas.1108174108}
  {\bibfield  {journal} {\bibinfo  {journal} {Proceedings of the National
  Academy of Sciences}\ }\textbf {\bibinfo {volume} {108}},\ \bibinfo {pages}
  {12233} (\bibinfo {year} {2011})}\BibitemShut {NoStop}%
\bibitem [{\citenamefont {Liu}\ \emph {et~al.}(2019)\citenamefont {Liu},
  \citenamefont {Liu},\ and\ \citenamefont {Dai}}]{Liu2019}%
  \BibitemOpen
  \bibfield  {author} {\bibinfo {author} {\bibfnamefont {J.}~\bibnamefont
  {Liu}}, \bibinfo {author} {\bibfnamefont {J.}~\bibnamefont {Liu}},\ and\
  \bibinfo {author} {\bibfnamefont {X.}~\bibnamefont {Dai}},\ }\bibfield
  {title} {\bibinfo {title} {{Pseudo Landau level representation of twisted
  bilayer graphene: Band topology and implications on the correlated insulating
  phase}},\ }\href {https://doi.org/10.1103/PhysRevB.99.155415} {\bibfield
  {journal} {\bibinfo  {journal} {Phys. Rev. B}\ }\textbf {\bibinfo {volume}
  {99}},\ \bibinfo {pages} {155415} (\bibinfo {year} {2019})}\BibitemShut
  {NoStop}%
\bibitem [{\citenamefont {Yuan}\ \emph {et~al.}(2019)\citenamefont {Yuan},
  \citenamefont {Isobe},\ and\ \citenamefont {Fu}}]{Yuan2019}%
  \BibitemOpen
  \bibfield  {author} {\bibinfo {author} {\bibfnamefont {N.~F.~Q.}\
  \bibnamefont {Yuan}}, \bibinfo {author} {\bibfnamefont {H.}~\bibnamefont
  {Isobe}},\ and\ \bibinfo {author} {\bibfnamefont {L.}~\bibnamefont {Fu}},\
  }\bibfield  {title} {\bibinfo {title} {{Magic of high-order van Hove
  singularity}},\ }\href {https://doi.org/10.1038/s41467-019-13670-9}
  {\bibfield  {journal} {\bibinfo  {journal} {Nature Communications}\ }\textbf
  {\bibinfo {volume} {10}},\ \bibinfo {pages} {5769} (\bibinfo {year}
  {2019})}\BibitemShut {NoStop}%
\bibitem [{\citenamefont {Song}\ \emph {et~al.}(2019)\citenamefont {Song},
  \citenamefont {Wang}, \citenamefont {Shi}, \citenamefont {Li}, \citenamefont
  {Fang},\ and\ \citenamefont {Bernevig}}]{Song2019}%
  \BibitemOpen
  \bibfield  {author} {\bibinfo {author} {\bibfnamefont {Z.}~\bibnamefont
  {Song}}, \bibinfo {author} {\bibfnamefont {Z.}~\bibnamefont {Wang}}, \bibinfo
  {author} {\bibfnamefont {W.}~\bibnamefont {Shi}}, \bibinfo {author}
  {\bibfnamefont {G.}~\bibnamefont {Li}}, \bibinfo {author} {\bibfnamefont
  {C.}~\bibnamefont {Fang}},\ and\ \bibinfo {author} {\bibfnamefont {B.~A.}\
  \bibnamefont {Bernevig}},\ }\bibfield  {title} {\bibinfo {title} {{All Magic
  Angles in Twisted Bilayer Graphene are Topological}},\ }\href
  {https://doi.org/10.1103/PhysRevLett.123.036401} {\bibfield  {journal}
  {\bibinfo  {journal} {Phys. Rev. Lett.}\ }\textbf {\bibinfo {volume} {123}},\
  \bibinfo {pages} {036401} (\bibinfo {year} {2019})}\BibitemShut {NoStop}%
\bibitem [{\citenamefont {Tarnopolsky}\ \emph {et~al.}(2019)\citenamefont
  {Tarnopolsky}, \citenamefont {Kruchkov},\ and\ \citenamefont
  {Vishwanath}}]{Tarnopolsky2019}%
  \BibitemOpen
  \bibfield  {author} {\bibinfo {author} {\bibfnamefont {G.}~\bibnamefont
  {Tarnopolsky}}, \bibinfo {author} {\bibfnamefont {A.~J.}\ \bibnamefont
  {Kruchkov}},\ and\ \bibinfo {author} {\bibfnamefont {A.}~\bibnamefont
  {Vishwanath}},\ }\bibfield  {title} {\bibinfo {title} {{Origin of Magic
  Angles in Twisted Bilayer Graphene}},\ }\href
  {https://doi.org/10.1103/PhysRevLett.122.106405} {\bibfield  {journal}
  {\bibinfo  {journal} {Phys. Rev. Lett.}\ }\textbf {\bibinfo {volume} {122}},\
  \bibinfo {pages} {106405} (\bibinfo {year} {2019})}\BibitemShut {NoStop}%
\bibitem [{\citenamefont {Kim}\ \emph {et~al.}(2017)\citenamefont {Kim},
  \citenamefont {DaSilva}, \citenamefont {Huang}, \citenamefont {Fallahazad},
  \citenamefont {Larentis}, \citenamefont {Taniguchi}, \citenamefont
  {Watanabe}, \citenamefont {LeRoy}, \citenamefont {MacDonald},\ and\
  \citenamefont {Tutuc}}]{Kim2017}%
  \BibitemOpen
  \bibfield  {author} {\bibinfo {author} {\bibfnamefont {K.}~\bibnamefont
  {Kim}}, \bibinfo {author} {\bibfnamefont {A.}~\bibnamefont {DaSilva}},
  \bibinfo {author} {\bibfnamefont {S.}~\bibnamefont {Huang}}, \bibinfo
  {author} {\bibfnamefont {B.}~\bibnamefont {Fallahazad}}, \bibinfo {author}
  {\bibfnamefont {S.}~\bibnamefont {Larentis}}, \bibinfo {author}
  {\bibfnamefont {T.}~\bibnamefont {Taniguchi}}, \bibinfo {author}
  {\bibfnamefont {K.}~\bibnamefont {Watanabe}}, \bibinfo {author}
  {\bibfnamefont {B.~J.}\ \bibnamefont {LeRoy}}, \bibinfo {author}
  {\bibfnamefont {A.~H.}\ \bibnamefont {MacDonald}},\ and\ \bibinfo {author}
  {\bibfnamefont {E.}~\bibnamefont {Tutuc}},\ }\bibfield  {title} {\bibinfo
  {title} {Tunable moir{\'{e}} bands and strong correlations in
  small-twist-angle bilayer graphene},\ }\href
  {https://doi.org/10.1073/pnas.1620140114} {\bibfield  {journal} {\bibinfo
  {journal} {Proceedings of the National Academy of Sciences}\ }\textbf
  {\bibinfo {volume} {114}},\ \bibinfo {pages} {3364} (\bibinfo {year}
  {2017})}\BibitemShut {NoStop}%
\bibitem [{\citenamefont {Ojaj\"arvi}\ \emph {et~al.}(2018)\citenamefont
  {Ojaj\"arvi}, \citenamefont {Hyart}, \citenamefont {Silaev},\ and\
  \citenamefont {Heikkil\"a}}]{Ojajarvi2018}%
  \BibitemOpen
  \bibfield  {author} {\bibinfo {author} {\bibfnamefont {R.}~\bibnamefont
  {Ojaj\"arvi}}, \bibinfo {author} {\bibfnamefont {T.}~\bibnamefont {Hyart}},
  \bibinfo {author} {\bibfnamefont {M.~A.}\ \bibnamefont {Silaev}},\ and\
  \bibinfo {author} {\bibfnamefont {T.~T.}\ \bibnamefont {Heikkil\"a}},\
  }\bibfield  {title} {\bibinfo {title} {{Competition of electron-phonon
  mediated superconductivity and Stoner magnetism on a flat band}},\ }\href
  {https://doi.org/10.1103/PhysRevB.98.054515} {\bibfield  {journal} {\bibinfo
  {journal} {Phys. Rev. B}\ }\textbf {\bibinfo {volume} {98}},\ \bibinfo
  {pages} {054515} (\bibinfo {year} {2018})}\BibitemShut {NoStop}%
\bibitem [{\citenamefont {Cao}\ \emph {et~al.}(2018{\natexlab{a}})\citenamefont
  {Cao}, \citenamefont {Fatemi}, \citenamefont {Fang}, \citenamefont
  {Watanabe}, \citenamefont {Taniguchi}, \citenamefont {Kaxiras},\ and\
  \citenamefont {Jarillo-Herrero}}]{Cao2018a}%
  \BibitemOpen
  \bibfield  {author} {\bibinfo {author} {\bibfnamefont {Y.}~\bibnamefont
  {Cao}}, \bibinfo {author} {\bibfnamefont {V.}~\bibnamefont {Fatemi}},
  \bibinfo {author} {\bibfnamefont {S.}~\bibnamefont {Fang}}, \bibinfo {author}
  {\bibfnamefont {K.}~\bibnamefont {Watanabe}}, \bibinfo {author}
  {\bibfnamefont {T.}~\bibnamefont {Taniguchi}}, \bibinfo {author}
  {\bibfnamefont {E.}~\bibnamefont {Kaxiras}},\ and\ \bibinfo {author}
  {\bibfnamefont {P.}~\bibnamefont {Jarillo-Herrero}},\ }\bibfield  {title}
  {\bibinfo {title} {{Unconventional superconductivity in magic-angle graphene
  superlattices}},\ }\href {https://doi.org/10.1038/nature26160} {\bibfield
  {journal} {\bibinfo  {journal} {Nature}\ }\textbf {\bibinfo {volume} {556}},\
  \bibinfo {pages} {43} (\bibinfo {year} {2018}{\natexlab{a}})}\BibitemShut
  {NoStop}%
\bibitem [{\citenamefont {Cao}\ \emph {et~al.}(2018{\natexlab{b}})\citenamefont
  {Cao}, \citenamefont {Fatemi}, \citenamefont {Demir}, \citenamefont {Fang},
  \citenamefont {Tomarken}, \citenamefont {Luo}, \citenamefont
  {Sanchez-Yamagishi}, \citenamefont {Watanabe}, \citenamefont {Taniguchi},
  \citenamefont {Kaxiras}, \citenamefont {Ashoori},\ and\ \citenamefont
  {Jarillo-Herrero}}]{Cao2018b}%
  \BibitemOpen
  \bibfield  {author} {\bibinfo {author} {\bibfnamefont {Y.}~\bibnamefont
  {Cao}}, \bibinfo {author} {\bibfnamefont {V.}~\bibnamefont {Fatemi}},
  \bibinfo {author} {\bibfnamefont {A.}~\bibnamefont {Demir}}, \bibinfo
  {author} {\bibfnamefont {S.}~\bibnamefont {Fang}}, \bibinfo {author}
  {\bibfnamefont {S.~L.}\ \bibnamefont {Tomarken}}, \bibinfo {author}
  {\bibfnamefont {J.~Y.}\ \bibnamefont {Luo}}, \bibinfo {author} {\bibfnamefont
  {J.~D.}\ \bibnamefont {Sanchez-Yamagishi}}, \bibinfo {author} {\bibfnamefont
  {K.}~\bibnamefont {Watanabe}}, \bibinfo {author} {\bibfnamefont
  {T.}~\bibnamefont {Taniguchi}}, \bibinfo {author} {\bibfnamefont
  {E.}~\bibnamefont {Kaxiras}}, \bibinfo {author} {\bibfnamefont {R.~C.}\
  \bibnamefont {Ashoori}},\ and\ \bibinfo {author} {\bibfnamefont
  {P.}~\bibnamefont {Jarillo-Herrero}},\ }\bibfield  {title} {\bibinfo {title}
  {Correlated insulator behaviour at half-filling in magic-angle graphene
  superlattices},\ }\href {https://doi.org/10.1038/nature26154} {\bibfield
  {journal} {\bibinfo  {journal} {Nature}\ }\textbf {\bibinfo {volume} {556}}
  (\bibinfo {year} {2018}{\natexlab{b}})}\BibitemShut {NoStop}%
\bibitem [{\citenamefont {Isobe}\ \emph {et~al.}(2018)\citenamefont {Isobe},
  \citenamefont {Yuan},\ and\ \citenamefont {Fu}}]{Liang2018}%
  \BibitemOpen
  \bibfield  {author} {\bibinfo {author} {\bibfnamefont {H.}~\bibnamefont
  {Isobe}}, \bibinfo {author} {\bibfnamefont {N.~F.~Q.}\ \bibnamefont {Yuan}},\
  and\ \bibinfo {author} {\bibfnamefont {L.}~\bibnamefont {Fu}},\ }\bibfield
  {title} {\bibinfo {title} {{Unconventional Superconductivity and Density
  Waves in Twisted Bilayer Graphene}},\ }\href
  {https://doi.org/10.1103/PhysRevX.8.041041} {\bibfield  {journal} {\bibinfo
  {journal} {Phys. Rev. X}\ }\textbf {\bibinfo {volume} {8}},\ \bibinfo {pages}
  {041041} (\bibinfo {year} {2018})}\BibitemShut {NoStop}%
\bibitem [{\citenamefont {Wu}\ \emph {et~al.}(2018)\citenamefont {Wu},
  \citenamefont {MacDonald},\ and\ \citenamefont {Martin}}]{Wu2018}%
  \BibitemOpen
  \bibfield  {author} {\bibinfo {author} {\bibfnamefont {F.}~\bibnamefont
  {Wu}}, \bibinfo {author} {\bibfnamefont {A.~H.}\ \bibnamefont {MacDonald}},\
  and\ \bibinfo {author} {\bibfnamefont {I.}~\bibnamefont {Martin}},\
  }\bibfield  {title} {\bibinfo {title} {{Theory of Phonon-Mediated
  Superconductivity in Twisted Bilayer Graphene}},\ }\href
  {https://doi.org/10.1103/PhysRevLett.121.257001} {\bibfield  {journal}
  {\bibinfo  {journal} {Physical Review Letters}\ }\textbf {\bibinfo {volume}
  {121}},\ \bibinfo {pages} {257001} (\bibinfo {year} {2018})}\BibitemShut
  {NoStop}%
\bibitem [{\citenamefont {Peltonen}\ \emph {et~al.}(2018)\citenamefont
  {Peltonen}, \citenamefont {Ojaj\"arvi},\ and\ \citenamefont
  {Heikkil\"a}}]{PeltonenPhysRevB.98.220504}%
  \BibitemOpen
  \bibfield  {author} {\bibinfo {author} {\bibfnamefont {T.~J.}\ \bibnamefont
  {Peltonen}}, \bibinfo {author} {\bibfnamefont {R.}~\bibnamefont
  {Ojaj\"arvi}},\ and\ \bibinfo {author} {\bibfnamefont {T.~T.}\ \bibnamefont
  {Heikkil\"a}},\ }\bibfield  {title} {\bibinfo {title} {Mean-field theory for
  superconductivity in twisted bilayer graphene},\ }\href
  {https://doi.org/10.1103/PhysRevB.98.220504} {\bibfield  {journal} {\bibinfo
  {journal} {Phys. Rev. B}\ }\textbf {\bibinfo {volume} {98}},\ \bibinfo
  {pages} {220504} (\bibinfo {year} {2018})}\BibitemShut {NoStop}%
\bibitem [{\citenamefont {Lu}\ \emph {et~al.}(2019)\citenamefont {Lu},
  \citenamefont {Stepanov}, \citenamefont {Yang}, \citenamefont {Xie},
  \citenamefont {Aamir}, \citenamefont {Das}, \citenamefont {Urgell},
  \citenamefont {Watanabe}, \citenamefont {Taniguchi}, \citenamefont {Zhang},
  \citenamefont {Bachtold}, \citenamefont {MacDonald},\ and\ \citenamefont
  {Efetov}}]{Lu2019}%
  \BibitemOpen
  \bibfield  {author} {\bibinfo {author} {\bibfnamefont {X.}~\bibnamefont
  {Lu}}, \bibinfo {author} {\bibfnamefont {P.}~\bibnamefont {Stepanov}},
  \bibinfo {author} {\bibfnamefont {W.}~\bibnamefont {Yang}}, \bibinfo {author}
  {\bibfnamefont {M.}~\bibnamefont {Xie}}, \bibinfo {author} {\bibfnamefont
  {M.~A.}\ \bibnamefont {Aamir}}, \bibinfo {author} {\bibfnamefont
  {I.}~\bibnamefont {Das}}, \bibinfo {author} {\bibfnamefont {C.}~\bibnamefont
  {Urgell}}, \bibinfo {author} {\bibfnamefont {K.}~\bibnamefont {Watanabe}},
  \bibinfo {author} {\bibfnamefont {T.}~\bibnamefont {Taniguchi}}, \bibinfo
  {author} {\bibfnamefont {G.}~\bibnamefont {Zhang}}, \bibinfo {author}
  {\bibfnamefont {A.}~\bibnamefont {Bachtold}}, \bibinfo {author}
  {\bibfnamefont {A.~H.}\ \bibnamefont {MacDonald}},\ and\ \bibinfo {author}
  {\bibfnamefont {D.~K.}\ \bibnamefont {Efetov}},\ }\bibfield  {title}
  {\bibinfo {title} {{Superconductors, orbital magnets and correlated states in
  magic-angle bilayer graphene}},\ }\href
  {https://doi.org/10.1038/s41586-019-1695-0} {\bibfield  {journal} {\bibinfo
  {journal} {Nature}\ }\textbf {\bibinfo {volume} {574}},\ \bibinfo {pages}
  {653} (\bibinfo {year} {2019})}\BibitemShut {NoStop}%
\bibitem [{\citenamefont {Kozii}\ \emph {et~al.}(2019)\citenamefont {Kozii},
  \citenamefont {Isobe}, \citenamefont {Venderbos},\ and\ \citenamefont
  {Fu}}]{Liang2019}%
  \BibitemOpen
  \bibfield  {author} {\bibinfo {author} {\bibfnamefont {V.}~\bibnamefont
  {Kozii}}, \bibinfo {author} {\bibfnamefont {H.}~\bibnamefont {Isobe}},
  \bibinfo {author} {\bibfnamefont {J.~W.~F.}\ \bibnamefont {Venderbos}},\ and\
  \bibinfo {author} {\bibfnamefont {L.}~\bibnamefont {Fu}},\ }\bibfield
  {title} {\bibinfo {title} {{Nematic superconductivity stabilized by density
  wave fluctuations: Possible application to twisted bilayer graphene}},\
  }\href {https://doi.org/10.1103/PhysRevB.99.144507} {\bibfield  {journal}
  {\bibinfo  {journal} {Phys. Rev. B}\ }\textbf {\bibinfo {volume} {99}},\
  \bibinfo {pages} {144507} (\bibinfo {year} {2019})}\BibitemShut {NoStop}%
\bibitem [{\citenamefont {Yankowitz}\ \emph {et~al.}(2019)\citenamefont
  {Yankowitz}, \citenamefont {Chen}, \citenamefont {Polshyn}, \citenamefont
  {Zhang}, \citenamefont {Watanabe}, \citenamefont {Taniguchi}, \citenamefont
  {Graf}, \citenamefont {Young},\ and\ \citenamefont {Dean}}]{Yankowitz2019}%
  \BibitemOpen
  \bibfield  {author} {\bibinfo {author} {\bibfnamefont {M.}~\bibnamefont
  {Yankowitz}}, \bibinfo {author} {\bibfnamefont {S.}~\bibnamefont {Chen}},
  \bibinfo {author} {\bibfnamefont {H.}~\bibnamefont {Polshyn}}, \bibinfo
  {author} {\bibfnamefont {Y.}~\bibnamefont {Zhang}}, \bibinfo {author}
  {\bibfnamefont {K.}~\bibnamefont {Watanabe}}, \bibinfo {author}
  {\bibfnamefont {T.}~\bibnamefont {Taniguchi}}, \bibinfo {author}
  {\bibfnamefont {D.}~\bibnamefont {Graf}}, \bibinfo {author} {\bibfnamefont
  {A.~F.}\ \bibnamefont {Young}},\ and\ \bibinfo {author} {\bibfnamefont
  {C.~R.}\ \bibnamefont {Dean}},\ }\bibfield  {title} {\bibinfo {title} {Tuning
  superconductivity in twisted bilayer graphene},\ }\href
  {https://doi.org/10.1126/science.aav1910} {\bibfield  {journal} {\bibinfo
  {journal} {Science}\ }\textbf {\bibinfo {volume} {363}},\ \bibinfo {pages}
  {1059} (\bibinfo {year} {2019})}\BibitemShut {NoStop}%
\bibitem [{\citenamefont {Hazra}\ \emph {et~al.}(2019)\citenamefont {Hazra},
  \citenamefont {Verma},\ and\ \citenamefont {Randeria}}]{Hazra2019}%
  \BibitemOpen
  \bibfield  {author} {\bibinfo {author} {\bibfnamefont {T.}~\bibnamefont
  {Hazra}}, \bibinfo {author} {\bibfnamefont {N.}~\bibnamefont {Verma}},\ and\
  \bibinfo {author} {\bibfnamefont {M.}~\bibnamefont {Randeria}},\ }\bibfield
  {title} {\bibinfo {title} {{Bounds on the Superconducting Transition
  Temperature : Applications to Twisted Bilayer Graphene and Cold Atoms}},\
  }\href {https://doi.org/10.1103/PhysRevX.9.031049} {\bibfield  {journal}
  {\bibinfo  {journal} {Physical Review X}\ }\textbf {\bibinfo {volume} {9}},\
  \bibinfo {pages} {31049} (\bibinfo {year} {2019})}\BibitemShut {NoStop}%
\bibitem [{\citenamefont {Hu}\ \emph {et~al.}(2019)\citenamefont {Hu},
  \citenamefont {Hyart}, \citenamefont {Pikulin},\ and\ \citenamefont
  {Rossi}}]{Hu2019}%
  \BibitemOpen
  \bibfield  {author} {\bibinfo {author} {\bibfnamefont {X.}~\bibnamefont
  {Hu}}, \bibinfo {author} {\bibfnamefont {T.}~\bibnamefont {Hyart}}, \bibinfo
  {author} {\bibfnamefont {D.~I.}\ \bibnamefont {Pikulin}},\ and\ \bibinfo
  {author} {\bibfnamefont {E.}~\bibnamefont {Rossi}},\ }\bibfield  {title}
  {\bibinfo {title} {Geometric and conventional contribution to the superfluid
  weight in twisted bilayer graphene},\ }\href
  {https://doi.org/10.1103/PhysRevLett.123.237002} {\bibfield  {journal}
  {\bibinfo  {journal} {Phys. Rev. Lett.}\ }\textbf {\bibinfo {volume} {123}},\
  \bibinfo {pages} {237002} (\bibinfo {year} {2019})}\BibitemShut {NoStop}%
\bibitem [{\citenamefont {Xie}\ \emph {et~al.}(2020)\citenamefont {Xie},
  \citenamefont {Song}, \citenamefont {Lian},\ and\ \citenamefont
  {Bernevig}}]{Fang2020}%
  \BibitemOpen
  \bibfield  {author} {\bibinfo {author} {\bibfnamefont {F.}~\bibnamefont
  {Xie}}, \bibinfo {author} {\bibfnamefont {Z.}~\bibnamefont {Song}}, \bibinfo
  {author} {\bibfnamefont {B.}~\bibnamefont {Lian}},\ and\ \bibinfo {author}
  {\bibfnamefont {B.~A.}\ \bibnamefont {Bernevig}},\ }\bibfield  {title}
  {\bibinfo {title} {{Topology-Bounded Superfluid Weight in Twisted Bilayer
  Graphene}},\ }\href {https://doi.org/10.1103/PhysRevLett.124.167002}
  {\bibfield  {journal} {\bibinfo  {journal} {Phys. Rev. Lett.}\ }\textbf
  {\bibinfo {volume} {124}},\ \bibinfo {pages} {167002} (\bibinfo {year}
  {2020})}\BibitemShut {NoStop}%
\bibitem [{\citenamefont {Julku}\ \emph {et~al.}(2020)\citenamefont {Julku},
  \citenamefont {Peltonen}, \citenamefont {Liang}, \citenamefont {Heikkil\"a},\
  and\ \citenamefont {T\"orm\"a}}]{Julku2020}%
  \BibitemOpen
  \bibfield  {author} {\bibinfo {author} {\bibfnamefont {A.}~\bibnamefont
  {Julku}}, \bibinfo {author} {\bibfnamefont {T.~J.}\ \bibnamefont {Peltonen}},
  \bibinfo {author} {\bibfnamefont {L.}~\bibnamefont {Liang}}, \bibinfo
  {author} {\bibfnamefont {T.~T.}\ \bibnamefont {Heikkil\"a}},\ and\ \bibinfo
  {author} {\bibfnamefont {P.}~\bibnamefont {T\"orm\"a}},\ }\bibfield  {title}
  {\bibinfo {title} {{Superfluid weight and Berezinskii-Kosterlitz-Thouless
  transition temperature of twisted bilayer graphene}},\ }\href
  {https://doi.org/10.1103/PhysRevB.101.060505} {\bibfield  {journal} {\bibinfo
   {journal} {Phys. Rev. B}\ }\textbf {\bibinfo {volume} {101}},\ \bibinfo
  {pages} {060505} (\bibinfo {year} {2020})}\BibitemShut {NoStop}%
\bibitem [{\citenamefont {Andrei}\ and\ \citenamefont
  {MacDonald}(2020)}]{Andrei2020}%
  \BibitemOpen
  \bibfield  {author} {\bibinfo {author} {\bibfnamefont {E.~Y.}\ \bibnamefont
  {Andrei}}\ and\ \bibinfo {author} {\bibfnamefont {A.~H.}\ \bibnamefont
  {MacDonald}},\ }\bibfield  {title} {\bibinfo {title} {Graphene bilayers with
  a twist},\ }\href {https://doi.org/10.1038/s41563-020-00840-0} {\bibfield
  {journal} {\bibinfo  {journal} {Nature Materials}\ }\textbf {\bibinfo
  {volume} {19}},\ \bibinfo {pages} {1265} (\bibinfo {year}
  {2020})}\BibitemShut {NoStop}%
\bibitem [{\citenamefont {Sanjuan~Ciepielewski}\ \emph
  {et~al.}(2022)\citenamefont {Sanjuan~Ciepielewski}, \citenamefont
  {Tworzyd\l{}o}, \citenamefont {Hyart},\ and\ \citenamefont
  {Lau}}]{Sanjuan2022}%
  \BibitemOpen
  \bibfield  {author} {\bibinfo {author} {\bibfnamefont {A.}~\bibnamefont
  {Sanjuan~Ciepielewski}}, \bibinfo {author} {\bibfnamefont {J.}~\bibnamefont
  {Tworzyd\l{}o}}, \bibinfo {author} {\bibfnamefont {T.}~\bibnamefont
  {Hyart}},\ and\ \bibinfo {author} {\bibfnamefont {A.}~\bibnamefont {Lau}},\
  }\bibfield  {title} {\bibinfo {title} {Transport signatures of van hove
  singularities in mesoscopic twisted bilayer graphene},\ }\href
  {https://doi.org/10.1103/PhysRevResearch.4.043145} {\bibfield  {journal}
  {\bibinfo  {journal} {Phys. Rev. Res.}\ }\textbf {\bibinfo {volume} {4}},\
  \bibinfo {pages} {043145} (\bibinfo {year} {2022})}\BibitemShut {NoStop}%
\bibitem [{\citenamefont {Uri}\ \emph {et~al.}(2020)\citenamefont {Uri},
  \citenamefont {Grover}, \citenamefont {Cao}, \citenamefont {Crosse},
  \citenamefont {Bagani}, \citenamefont {Rodan-Legrain}, \citenamefont
  {Myasoedov}, \citenamefont {Watanabe}, \citenamefont {Taniguchi},
  \citenamefont {Moon}, \citenamefont {Koshino}, \citenamefont
  {Jarillo-Herrero},\ and\ \citenamefont {Zeldov}}]{Uri2020}%
  \BibitemOpen
  \bibfield  {author} {\bibinfo {author} {\bibfnamefont {A.}~\bibnamefont
  {Uri}}, \bibinfo {author} {\bibfnamefont {S.}~\bibnamefont {Grover}},
  \bibinfo {author} {\bibfnamefont {Y.}~\bibnamefont {Cao}}, \bibinfo {author}
  {\bibfnamefont {J.~A.}\ \bibnamefont {Crosse}}, \bibinfo {author}
  {\bibfnamefont {K.}~\bibnamefont {Bagani}}, \bibinfo {author} {\bibfnamefont
  {D.}~\bibnamefont {Rodan-Legrain}}, \bibinfo {author} {\bibfnamefont
  {Y.}~\bibnamefont {Myasoedov}}, \bibinfo {author} {\bibfnamefont
  {K.}~\bibnamefont {Watanabe}}, \bibinfo {author} {\bibfnamefont
  {T.}~\bibnamefont {Taniguchi}}, \bibinfo {author} {\bibfnamefont
  {P.}~\bibnamefont {Moon}}, \bibinfo {author} {\bibfnamefont {M.}~\bibnamefont
  {Koshino}}, \bibinfo {author} {\bibfnamefont {P.}~\bibnamefont
  {Jarillo-Herrero}},\ and\ \bibinfo {author} {\bibfnamefont {E.}~\bibnamefont
  {Zeldov}},\ }\bibfield  {title} {\bibinfo {title} {Mapping the twist-angle
  disorder and landau levels in magic-angle graphene},\ }\href
  {https://doi.org/10.1038/s41586-020-2255-3} {\bibfield  {journal} {\bibinfo
  {journal} {Nature}\ }\textbf {\bibinfo {volume} {581}},\ \bibinfo {pages}
  {47} (\bibinfo {year} {2020})}\BibitemShut {NoStop}%
\bibitem [{\citenamefont {de~Jong}\ \emph {et~al.}(2022)\citenamefont
  {de~Jong}, \citenamefont {Benschop}, \citenamefont {Chen}, \citenamefont
  {Krasovskii}, \citenamefont {de~Dood}, \citenamefont {Tromp}, \citenamefont
  {Allan},\ and\ \citenamefont {van~der Molen}}]{deJong2022}%
  \BibitemOpen
  \bibfield  {author} {\bibinfo {author} {\bibfnamefont {T.~A.}\ \bibnamefont
  {de~Jong}}, \bibinfo {author} {\bibfnamefont {T.}~\bibnamefont {Benschop}},
  \bibinfo {author} {\bibfnamefont {X.}~\bibnamefont {Chen}}, \bibinfo {author}
  {\bibfnamefont {E.~E.}\ \bibnamefont {Krasovskii}}, \bibinfo {author}
  {\bibfnamefont {M.~J.~A.}\ \bibnamefont {de~Dood}}, \bibinfo {author}
  {\bibfnamefont {R.~M.}\ \bibnamefont {Tromp}}, \bibinfo {author}
  {\bibfnamefont {M.~P.}\ \bibnamefont {Allan}},\ and\ \bibinfo {author}
  {\bibfnamefont {S.~J.}\ \bibnamefont {van~der Molen}},\ }\bibfield  {title}
  {\bibinfo {title} {Imaging moir{\'e} deformation and dynamics in twisted
  bilayer graphene},\ }\href {https://doi.org/10.1038/s41467-021-27646-1}
  {\bibfield  {journal} {\bibinfo  {journal} {Nature Communications}\ }\textbf
  {\bibinfo {volume} {13}},\ \bibinfo {pages} {70} (\bibinfo {year}
  {2022})}\BibitemShut {NoStop}%
\bibitem [{\citenamefont {Hu}\ \emph {et~al.}(2024)\citenamefont {Hu},
  \citenamefont {Zhu}, \citenamefont {Hu}, \citenamefont {Wang}, \citenamefont
  {Shen}, \citenamefont {Yang}, \citenamefont {Zhu}, \citenamefont {Huan},
  \citenamefont {Xu},\ and\ \citenamefont {Gao}}]{Hu_2024}%
  \BibitemOpen
  \bibfield  {author} {\bibinfo {author} {\bibfnamefont {J.}~\bibnamefont
  {Hu}}, \bibinfo {author} {\bibfnamefont {S.}~\bibnamefont {Zhu}}, \bibinfo
  {author} {\bibfnamefont {Q.}~\bibnamefont {Hu}}, \bibinfo {author}
  {\bibfnamefont {Y.}~\bibnamefont {Wang}}, \bibinfo {author} {\bibfnamefont
  {C.}~\bibnamefont {Shen}}, \bibinfo {author} {\bibfnamefont {H.}~\bibnamefont
  {Yang}}, \bibinfo {author} {\bibfnamefont {X.}~\bibnamefont {Zhu}}, \bibinfo
  {author} {\bibfnamefont {Q.}~\bibnamefont {Huan}}, \bibinfo {author}
  {\bibfnamefont {Y.}~\bibnamefont {Xu}},\ and\ \bibinfo {author}
  {\bibfnamefont {H.-J.}\ \bibnamefont {Gao}},\ }\bibfield  {title} {\bibinfo
  {title} {Visualizing the local twist angle variation within and between
  domains of twisted bilayer graphene},\ }\href
  {https://doi.org/10.1088/0256-307X/41/3/037401} {\bibfield  {journal}
  {\bibinfo  {journal} {Chinese Physics Letters}\ }\textbf {\bibinfo {volume}
  {41}},\ \bibinfo {pages} {037401} (\bibinfo {year} {2024})}\BibitemShut
  {NoStop}%
\bibitem [{\citenamefont {Wilson}\ \emph {et~al.}(2020)\citenamefont {Wilson},
  \citenamefont {Fu}, \citenamefont {Das~Sarma},\ and\ \citenamefont
  {Pixley}}]{Wilson2020}%
  \BibitemOpen
  \bibfield  {author} {\bibinfo {author} {\bibfnamefont {J.~H.}\ \bibnamefont
  {Wilson}}, \bibinfo {author} {\bibfnamefont {Y.}~\bibnamefont {Fu}}, \bibinfo
  {author} {\bibfnamefont {S.}~\bibnamefont {Das~Sarma}},\ and\ \bibinfo
  {author} {\bibfnamefont {J.~H.}\ \bibnamefont {Pixley}},\ }\bibfield  {title}
  {\bibinfo {title} {Disorder in twisted bilayer graphene},\ }\href
  {https://doi.org/10.1103/PhysRevResearch.2.023325} {\bibfield  {journal}
  {\bibinfo  {journal} {Phys. Rev. Res.}\ }\textbf {\bibinfo {volume} {2}},\
  \bibinfo {pages} {023325} (\bibinfo {year} {2020})}\BibitemShut {NoStop}%
\bibitem [{\citenamefont {Sainz-Cruz}\ \emph {et~al.}(2021)\citenamefont
  {Sainz-Cruz}, \citenamefont {Cea}, \citenamefont {Pantale\'on},\ and\
  \citenamefont {Guinea}}]{Cruz2021}%
  \BibitemOpen
  \bibfield  {author} {\bibinfo {author} {\bibfnamefont {H.}~\bibnamefont
  {Sainz-Cruz}}, \bibinfo {author} {\bibfnamefont {T.}~\bibnamefont {Cea}},
  \bibinfo {author} {\bibfnamefont {P.~A.}\ \bibnamefont {Pantale\'on}},\ and\
  \bibinfo {author} {\bibfnamefont {F.}~\bibnamefont {Guinea}},\ }\bibfield
  {title} {\bibinfo {title} {High transmission in twisted bilayer graphene with
  angle disorder},\ }\href {https://doi.org/10.1103/PhysRevB.104.075144}
  {\bibfield  {journal} {\bibinfo  {journal} {Phys. Rev. B}\ }\textbf {\bibinfo
  {volume} {104}},\ \bibinfo {pages} {075144} (\bibinfo {year}
  {2021})}\BibitemShut {NoStop}%
\bibitem [{\citenamefont {Nakatsuji}\ and\ \citenamefont
  {Koshino}(2022)}]{Naoto2022}%
  \BibitemOpen
  \bibfield  {author} {\bibinfo {author} {\bibfnamefont {N.}~\bibnamefont
  {Nakatsuji}}\ and\ \bibinfo {author} {\bibfnamefont {M.}~\bibnamefont
  {Koshino}},\ }\bibfield  {title} {\bibinfo {title} {Moir\'e disorder effect
  in twisted bilayer graphene},\ }\href
  {https://doi.org/10.1103/PhysRevB.105.245408} {\bibfield  {journal} {\bibinfo
   {journal} {Phys. Rev. B}\ }\textbf {\bibinfo {volume} {105}},\ \bibinfo
  {pages} {245408} (\bibinfo {year} {2022})}\BibitemShut {NoStop}%
\bibitem [{\citenamefont {Padhi}\ \emph {et~al.}(2020)\citenamefont {Padhi},
  \citenamefont {Tiwari}, \citenamefont {Neupert},\ and\ \citenamefont
  {Ryu}}]{Padhi20}%
  \BibitemOpen
  \bibfield  {author} {\bibinfo {author} {\bibfnamefont {B.}~\bibnamefont
  {Padhi}}, \bibinfo {author} {\bibfnamefont {A.}~\bibnamefont {Tiwari}},
  \bibinfo {author} {\bibfnamefont {T.}~\bibnamefont {Neupert}},\ and\ \bibinfo
  {author} {\bibfnamefont {S.}~\bibnamefont {Ryu}},\ }\bibfield  {title}
  {\bibinfo {title} {{Transport across twist angle domains in moir\'e
  graphene}},\ }\href {https://doi.org/10.1103/PhysRevResearch.2.033458}
  {\bibfield  {journal} {\bibinfo  {journal} {Phys. Rev. Research}\ }\textbf
  {\bibinfo {volume} {2}},\ \bibinfo {pages} {033458} (\bibinfo {year}
  {2020})}\BibitemShut {NoStop}%
\bibitem [{\citenamefont {Joy}\ \emph {et~al.}(2020)\citenamefont {Joy},
  \citenamefont {Khalid},\ and\ \citenamefont {Skinner}}]{Joy2020}%
  \BibitemOpen
  \bibfield  {author} {\bibinfo {author} {\bibfnamefont {S.}~\bibnamefont
  {Joy}}, \bibinfo {author} {\bibfnamefont {S.}~\bibnamefont {Khalid}},\ and\
  \bibinfo {author} {\bibfnamefont {B.}~\bibnamefont {Skinner}},\ }\bibfield
  {title} {\bibinfo {title} {Transparent mirror effect in
  twist-angle-disordered bilayer graphene},\ }\href
  {https://doi.org/10.1103/PhysRevResearch.2.043416} {\bibfield  {journal}
  {\bibinfo  {journal} {Phys. Rev. Res.}\ }\textbf {\bibinfo {volume} {2}},\
  \bibinfo {pages} {043416} (\bibinfo {year} {2020})}\BibitemShut {NoStop}%
\bibitem [{\citenamefont {Thomson}\ and\ \citenamefont
  {Alicea}(2021)}]{Thomson2021}%
  \BibitemOpen
  \bibfield  {author} {\bibinfo {author} {\bibfnamefont {A.}~\bibnamefont
  {Thomson}}\ and\ \bibinfo {author} {\bibfnamefont {J.}~\bibnamefont
  {Alicea}},\ }\bibfield  {title} {\bibinfo {title} {Recovery of massless dirac
  fermions at charge neutrality in strongly interacting twisted bilayer
  graphene with disorder},\ }\href
  {https://doi.org/10.1103/PhysRevB.103.125138} {\bibfield  {journal} {\bibinfo
   {journal} {Phys. Rev. B}\ }\textbf {\bibinfo {volume} {103}},\ \bibinfo
  {pages} {125138} (\bibinfo {year} {2021})}\BibitemShut {NoStop}%
\bibitem [{\citenamefont {Wu}\ \emph {et~al.}(2019)\citenamefont {Wu},
  \citenamefont {Hwang},\ and\ \citenamefont {Das~Sarma}}]{WudasSarma19}%
  \BibitemOpen
  \bibfield  {author} {\bibinfo {author} {\bibfnamefont {F.}~\bibnamefont
  {Wu}}, \bibinfo {author} {\bibfnamefont {E.}~\bibnamefont {Hwang}},\ and\
  \bibinfo {author} {\bibfnamefont {S.}~\bibnamefont {Das~Sarma}},\ }\bibfield
  {title} {\bibinfo {title} {{Phonon-induced giant linear-in-$T$ resistivity in
  magic angle twisted bilayer graphene: Ordinary strangeness and exotic
  superconductivity}},\ }\href {https://doi.org/10.1103/PhysRevB.99.165112}
  {\bibfield  {journal} {\bibinfo  {journal} {Phys. Rev. B}\ }\textbf {\bibinfo
  {volume} {99}},\ \bibinfo {pages} {165112} (\bibinfo {year}
  {2019})}\BibitemShut {NoStop}%
\bibitem [{\citenamefont {Hwang}\ and\ \citenamefont
  {Das~Sarma}(2020)}]{HwangdasSarma}%
  \BibitemOpen
  \bibfield  {author} {\bibinfo {author} {\bibfnamefont {E.~H.}\ \bibnamefont
  {Hwang}}\ and\ \bibinfo {author} {\bibfnamefont {S.}~\bibnamefont
  {Das~Sarma}},\ }\bibfield  {title} {\bibinfo {title}
  {{Impurity-scattering-induced carrier transport in twisted bilayer
  graphene}},\ }\href {https://doi.org/10.1103/PhysRevResearch.2.013342}
  {\bibfield  {journal} {\bibinfo  {journal} {Phys. Rev. Research}\ }\textbf
  {\bibinfo {volume} {2}},\ \bibinfo {pages} {013342} (\bibinfo {year}
  {2020})}\BibitemShut {NoStop}%
\bibitem [{\citenamefont {Andelkovi\ifmmode~\acute{c}\else \'{c}\fi{}}\ \emph
  {et~al.}(2018)\citenamefont {Andelkovi\ifmmode~\acute{c}\else \'{c}\fi{}},
  \citenamefont {Covaci},\ and\ \citenamefont {Peeters}}]{Andelkovic2018}%
  \BibitemOpen
  \bibfield  {author} {\bibinfo {author} {\bibfnamefont {M.}~\bibnamefont
  {Andelkovi\ifmmode~\acute{c}\else \'{c}\fi{}}}, \bibinfo {author}
  {\bibfnamefont {L.}~\bibnamefont {Covaci}},\ and\ \bibinfo {author}
  {\bibfnamefont {F.~M.}\ \bibnamefont {Peeters}},\ }\bibfield  {title}
  {\bibinfo {title} {{DC conductivity of twisted bilayer graphene:
  Angle-dependent transport properties and effects of disorder}},\ }\href
  {https://doi.org/10.1103/PhysRevMaterials.2.034004} {\bibfield  {journal}
  {\bibinfo  {journal} {Phys. Rev. Materials}\ }\textbf {\bibinfo {volume}
  {2}},\ \bibinfo {pages} {034004} (\bibinfo {year} {2018})}\BibitemShut
  {NoStop}%
\bibitem [{\citenamefont {Hou}\ \emph {et~al.}(2024)\citenamefont {Hou},
  \citenamefont {Hu},\ and\ \citenamefont {Yang}}]{Hou2024}%
  \BibitemOpen
  \bibfield  {author} {\bibinfo {author} {\bibfnamefont {Z.}~\bibnamefont
  {Hou}}, \bibinfo {author} {\bibfnamefont {Y.-Y.}\ \bibnamefont {Hu}},\ and\
  \bibinfo {author} {\bibfnamefont {G.-W.}\ \bibnamefont {Yang}},\ }\bibfield
  {title} {\bibinfo {title} {Moir\'e pattern assisted commensuration resonance
  in disordered twisted bilayer graphene},\ }\href
  {https://doi.org/10.1103/PhysRevB.109.085412} {\bibfield  {journal} {\bibinfo
   {journal} {Phys. Rev. B}\ }\textbf {\bibinfo {volume} {109}},\ \bibinfo
  {pages} {085412} (\bibinfo {year} {2024})}\BibitemShut {NoStop}%
\bibitem [{\citenamefont {Bahamon}\ \emph {et~al.}(2020)\citenamefont
  {Bahamon}, \citenamefont {G{\'{o}}mez-Santos},\ and\ \citenamefont
  {Stauber}}]{Bahamon2020}%
  \BibitemOpen
  \bibfield  {author} {\bibinfo {author} {\bibfnamefont {D.~A.}\ \bibnamefont
  {Bahamon}}, \bibinfo {author} {\bibfnamefont {G.}~\bibnamefont
  {G{\'{o}}mez-Santos}},\ and\ \bibinfo {author} {\bibfnamefont
  {T.}~\bibnamefont {Stauber}},\ }\bibfield  {title} {\bibinfo {title}
  {Emergent magnetic texture in driven twisted bilayer graphene},\ }\href
  {https://doi.org/10.1039/d0nr02786c} {\bibfield  {journal} {\bibinfo
  {journal} {Nanoscale}\ }\textbf {\bibinfo {volume} {12}},\ \bibinfo {pages}
  {15383} (\bibinfo {year} {2020})}\BibitemShut {NoStop}%
\bibitem [{\citenamefont {Zhou}\ \emph {et~al.}(2010)\citenamefont {Zhou},
  \citenamefont {Liao}, \citenamefont {Zhou}, \citenamefont {Chen},\ and\
  \citenamefont {Zhou}}]{Zhou2010}%
  \BibitemOpen
  \bibfield  {author} {\bibinfo {author} {\bibfnamefont {B.~H.}\ \bibnamefont
  {Zhou}}, \bibinfo {author} {\bibfnamefont {W.~H.}\ \bibnamefont {Liao}},
  \bibinfo {author} {\bibfnamefont {B.~L.}\ \bibnamefont {Zhou}}, \bibinfo
  {author} {\bibfnamefont {K.-Q.}\ \bibnamefont {Chen}},\ and\ \bibinfo
  {author} {\bibfnamefont {G.~H.}\ \bibnamefont {Zhou}},\ }\bibfield  {title}
  {\bibinfo {title} {Electronic transport for a crossed graphene nanoribbon
  junction with and without doping},\ }\href
  {https://doi.org/10.1140/epjb/e2010-00181-7} {\bibfield  {journal} {\bibinfo
  {journal} {The European Physical Journal B}\ }\textbf {\bibinfo {volume}
  {76}},\ \bibinfo {pages} {421} (\bibinfo {year} {2010})}\BibitemShut
  {NoStop}%
\bibitem [{\citenamefont {Brandimarte}\ \emph {et~al.}(2017)\citenamefont
  {Brandimarte}, \citenamefont {Engelund}, \citenamefont {Papior},
  \citenamefont {Garcia-Lekue}, \citenamefont {Frederiksen},\ and\
  \citenamefont {S{\'{a}}nchez-Portal}}]{Brandimarte2017}%
  \BibitemOpen
  \bibfield  {author} {\bibinfo {author} {\bibfnamefont {P.}~\bibnamefont
  {Brandimarte}}, \bibinfo {author} {\bibfnamefont {M.}~\bibnamefont
  {Engelund}}, \bibinfo {author} {\bibfnamefont {N.}~\bibnamefont {Papior}},
  \bibinfo {author} {\bibfnamefont {A.}~\bibnamefont {Garcia-Lekue}}, \bibinfo
  {author} {\bibfnamefont {T.}~\bibnamefont {Frederiksen}},\ and\ \bibinfo
  {author} {\bibfnamefont {D.}~\bibnamefont {S{\'{a}}nchez-Portal}},\
  }\bibfield  {title} {\bibinfo {title} {A tunable electronic beam splitter
  realized with crossed graphene nanoribbons},\ }\href
  {https://doi.org/10.1063/1.4974895} {\bibfield  {journal} {\bibinfo
  {journal} {The Journal of Chemical Physics}\ }\textbf {\bibinfo {volume}
  {146}},\ \bibinfo {pages} {092318} (\bibinfo {year} {2017})}\BibitemShut
  {NoStop}%
\bibitem [{\citenamefont {Sanz}\ \emph {et~al.}(2020)\citenamefont {Sanz},
  \citenamefont {Brandimarte}, \citenamefont {Giedke}, \citenamefont
  {S\'anchez-Portal},\ and\ \citenamefont {Frederiksen}}]{Sanz2020}%
  \BibitemOpen
  \bibfield  {author} {\bibinfo {author} {\bibfnamefont {S.}~\bibnamefont
  {Sanz}}, \bibinfo {author} {\bibfnamefont {P.}~\bibnamefont {Brandimarte}},
  \bibinfo {author} {\bibfnamefont {G.}~\bibnamefont {Giedke}}, \bibinfo
  {author} {\bibfnamefont {D.}~\bibnamefont {S\'anchez-Portal}},\ and\ \bibinfo
  {author} {\bibfnamefont {T.}~\bibnamefont {Frederiksen}},\ }\bibfield
  {title} {\bibinfo {title} {{Crossed graphene nanoribbons as beam splitters
  and mirrors for electron quantum optics}},\ }\href
  {https://doi.org/10.1103/PhysRevB.102.035436} {\bibfield  {journal} {\bibinfo
   {journal} {Phys. Rev. B}\ }\textbf {\bibinfo {volume} {102}},\ \bibinfo
  {pages} {035436} (\bibinfo {year} {2020})}\BibitemShut {NoStop}%
\bibitem [{\citenamefont {Olyaei}\ \emph {et~al.}(2020)\citenamefont {Olyaei},
  \citenamefont {Amorim}, \citenamefont {Ribeiro},\ and\ \citenamefont
  {Castro}}]{Olyaei2020}%
  \BibitemOpen
  \bibfield  {author} {\bibinfo {author} {\bibfnamefont {H.~Z.}\ \bibnamefont
  {Olyaei}}, \bibinfo {author} {\bibfnamefont {B.}~\bibnamefont {Amorim}},
  \bibinfo {author} {\bibfnamefont {P.}~\bibnamefont {Ribeiro}},\ and\ \bibinfo
  {author} {\bibfnamefont {E.~V.}\ \bibnamefont {Castro}},\ }\href
  {https://doi.org/10.48550/ARXIV.2007.14498} {\bibinfo {title} {Ballistic
  charge transport in twisted bilayer graphene}} (\bibinfo {year}
  {2020})\BibitemShut {NoStop}%
\bibitem [{\citenamefont {Lopes~dos Santos}\ \emph {et~al.}(2007)\citenamefont
  {Lopes~dos Santos}, \citenamefont {Peres},\ and\ \citenamefont
  {Castro~Neto}}]{dosSantos2007}%
  \BibitemOpen
  \bibfield  {author} {\bibinfo {author} {\bibfnamefont {J.~M.~B.}\
  \bibnamefont {Lopes~dos Santos}}, \bibinfo {author} {\bibfnamefont
  {N.~M.~R.}\ \bibnamefont {Peres}},\ and\ \bibinfo {author} {\bibfnamefont
  {A.~H.}\ \bibnamefont {Castro~Neto}},\ }\bibfield  {title} {\bibinfo {title}
  {{Graphene Bilayer with a Twist: Electronic Structure}},\ }\href
  {https://doi.org/10.1103/PhysRevLett.99.256802} {\bibfield  {journal}
  {\bibinfo  {journal} {Phys. Rev. Lett.}\ }\textbf {\bibinfo {volume} {99}},\
  \bibinfo {pages} {256802} (\bibinfo {year} {2007})}\BibitemShut {NoStop}%
\bibitem [{\citenamefont {Mele}(2010)}]{Mele2010}%
  \BibitemOpen
  \bibfield  {author} {\bibinfo {author} {\bibfnamefont {E.~J.}\ \bibnamefont
  {Mele}},\ }\bibfield  {title} {\bibinfo {title} {Commensuration and
  interlayer coherence in twisted bilayer graphene},\ }\href
  {https://doi.org/10.1103/PhysRevB.81.161405} {\bibfield  {journal} {\bibinfo
  {journal} {Phys. Rev. B}\ }\textbf {\bibinfo {volume} {81}},\ \bibinfo
  {pages} {161405} (\bibinfo {year} {2010})}\BibitemShut {NoStop}%
\bibitem [{\citenamefont {Cao}\ \emph {et~al.}(2021)\citenamefont {Cao},
  \citenamefont {Wang}, \citenamefont {Qian}, \citenamefont {Liu},\ and\
  \citenamefont {Yao}}]{Cao2021}%
  \BibitemOpen
  \bibfield  {author} {\bibinfo {author} {\bibfnamefont {J.}~\bibnamefont
  {Cao}}, \bibinfo {author} {\bibfnamefont {M.}~\bibnamefont {Wang}}, \bibinfo
  {author} {\bibfnamefont {S.-F.}\ \bibnamefont {Qian}}, \bibinfo {author}
  {\bibfnamefont {C.-C.}\ \bibnamefont {Liu}},\ and\ \bibinfo {author}
  {\bibfnamefont {Y.}~\bibnamefont {Yao}},\ }\bibfield  {title} {\bibinfo
  {title} {{Ab initio four-band Wannier tight-binding model for generic twisted
  graphene systems}},\ }\href {https://doi.org/10.1103/PhysRevB.104.L081403}
  {\bibfield  {journal} {\bibinfo  {journal} {Phys. Rev. B}\ }\textbf {\bibinfo
  {volume} {104}},\ \bibinfo {pages} {L081403} (\bibinfo {year}
  {2021})}\BibitemShut {NoStop}%
\bibitem [{\citenamefont {Lin}\ and\ \citenamefont
  {Tom\'anek}(2018)}]{Lin2018}%
  \BibitemOpen
  \bibfield  {author} {\bibinfo {author} {\bibfnamefont {X.}~\bibnamefont
  {Lin}}\ and\ \bibinfo {author} {\bibfnamefont {D.}~\bibnamefont
  {Tom\'anek}},\ }\bibfield  {title} {\bibinfo {title} {Minimum model for the
  electronic structure of twisted bilayer graphene and related structures},\
  }\href {https://doi.org/10.1103/PhysRevB.98.081410} {\bibfield  {journal}
  {\bibinfo  {journal} {Phys. Rev. B}\ }\textbf {\bibinfo {volume} {98}},\
  \bibinfo {pages} {081410} (\bibinfo {year} {2018})}\BibitemShut {NoStop}%
\bibitem [{\citenamefont {Moon}\ and\ \citenamefont
  {Koshino}(2012)}]{Moon2012}%
  \BibitemOpen
  \bibfield  {author} {\bibinfo {author} {\bibfnamefont {P.}~\bibnamefont
  {Moon}}\ and\ \bibinfo {author} {\bibfnamefont {M.}~\bibnamefont {Koshino}},\
  }\bibfield  {title} {\bibinfo {title} {{Energy spectrum and quantum Hall
  effect in twisted bilayer graphene}},\ }\href
  {https://doi.org/10.1103/PhysRevB.85.195458} {\bibfield  {journal} {\bibinfo
  {journal} {Phys. Rev. B}\ }\textbf {\bibinfo {volume} {85}},\ \bibinfo
  {pages} {195458} (\bibinfo {year} {2012})}\BibitemShut {NoStop}%
\bibitem [{\citenamefont {Po}\ \emph {et~al.}(2019)\citenamefont {Po},
  \citenamefont {Zou}, \citenamefont {Senthil},\ and\ \citenamefont
  {Vishwanath}}]{Po2019}%
  \BibitemOpen
  \bibfield  {author} {\bibinfo {author} {\bibfnamefont {H.~C.}\ \bibnamefont
  {Po}}, \bibinfo {author} {\bibfnamefont {L.}~\bibnamefont {Zou}}, \bibinfo
  {author} {\bibfnamefont {T.}~\bibnamefont {Senthil}},\ and\ \bibinfo {author}
  {\bibfnamefont {A.}~\bibnamefont {Vishwanath}},\ }\bibfield  {title}
  {\bibinfo {title} {Faithful tight-binding models and fragile topology of
  magic-angle bilayer graphene},\ }\href
  {https://doi.org/10.1103/PhysRevB.99.195455} {\bibfield  {journal} {\bibinfo
  {journal} {Phys. Rev. B}\ }\textbf {\bibinfo {volume} {99}},\ \bibinfo
  {pages} {195455} (\bibinfo {year} {2019})}\BibitemShut {NoStop}%
\bibitem [{\citenamefont {Neto}\ \emph {et~al.}(2009)\citenamefont {Neto},
  \citenamefont {Guinea}, \citenamefont {Peres}, \citenamefont {Novoselov},\
  and\ \citenamefont {Geim}}]{CastroNeto2009}%
  \BibitemOpen
  \bibfield  {author} {\bibinfo {author} {\bibfnamefont {A.~H.~C.}\
  \bibnamefont {Neto}}, \bibinfo {author} {\bibfnamefont {F.}~\bibnamefont
  {Guinea}}, \bibinfo {author} {\bibfnamefont {N.~M.~R.}\ \bibnamefont
  {Peres}}, \bibinfo {author} {\bibfnamefont {K.~S.}\ \bibnamefont
  {Novoselov}},\ and\ \bibinfo {author} {\bibfnamefont {A.~K.}\ \bibnamefont
  {Geim}},\ }\bibfield  {title} {\bibinfo {title} {The electronic properties of
  graphene},\ }\href {https://doi.org/10.1103/revmodphys.81.109} {\bibfield
  {journal} {\bibinfo  {journal} {Reviews of Modern Physics}\ }\textbf
  {\bibinfo {volume} {81}},\ \bibinfo {pages} {109} (\bibinfo {year}
  {2009})}\BibitemShut {NoStop}%
\bibitem [{\citenamefont {Groth}\ \emph {et~al.}(2014)\citenamefont {Groth},
  \citenamefont {Wimmer}, \citenamefont {Akhmerov},\ and\ \citenamefont
  {Waintal}}]{Groth2014}%
  \BibitemOpen
  \bibfield  {author} {\bibinfo {author} {\bibfnamefont {C.~W.}\ \bibnamefont
  {Groth}}, \bibinfo {author} {\bibfnamefont {M.}~\bibnamefont {Wimmer}},
  \bibinfo {author} {\bibfnamefont {A.~R.}\ \bibnamefont {Akhmerov}},\ and\
  \bibinfo {author} {\bibfnamefont {X.}~\bibnamefont {Waintal}},\ }\bibfield
  {title} {\bibinfo {title} {Kwant: a software package for quantum transport},\
  }\href {https://doi.org/10.1088/1367-2630/16/6/063065} {\bibfield  {journal}
  {\bibinfo  {journal} {New Journal of Physics}\ }\textbf {\bibinfo {volume}
  {16}},\ \bibinfo {pages} {063065} (\bibinfo {year} {2014})}\BibitemShut
  {NoStop}%
\bibitem [{\citenamefont {Veyrat}\ \emph {et~al.}(2020)\citenamefont {Veyrat},
  \citenamefont {D{\'{e}}prez}, \citenamefont {Coissard}, \citenamefont {Li},
  \citenamefont {Gay}, \citenamefont {Watanabe}, \citenamefont {Taniguchi},
  \citenamefont {Han}, \citenamefont {Piot}, \citenamefont {Sellier},\ and\
  \citenamefont {Sac{\'{e}}p{\'{e}}}}]{Veyrat2020}%
  \BibitemOpen
  \bibfield  {author} {\bibinfo {author} {\bibfnamefont {L.}~\bibnamefont
  {Veyrat}}, \bibinfo {author} {\bibfnamefont {C.}~\bibnamefont
  {D{\'{e}}prez}}, \bibinfo {author} {\bibfnamefont {A.}~\bibnamefont
  {Coissard}}, \bibinfo {author} {\bibfnamefont {X.}~\bibnamefont {Li}},
  \bibinfo {author} {\bibfnamefont {F.}~\bibnamefont {Gay}}, \bibinfo {author}
  {\bibfnamefont {K.}~\bibnamefont {Watanabe}}, \bibinfo {author}
  {\bibfnamefont {T.}~\bibnamefont {Taniguchi}}, \bibinfo {author}
  {\bibfnamefont {Z.}~\bibnamefont {Han}}, \bibinfo {author} {\bibfnamefont
  {B.~A.}\ \bibnamefont {Piot}}, \bibinfo {author} {\bibfnamefont
  {H.}~\bibnamefont {Sellier}},\ and\ \bibinfo {author} {\bibfnamefont
  {B.}~\bibnamefont {Sac{\'{e}}p{\'{e}}}},\ }\bibfield  {title} {\bibinfo
  {title} {{Helical quantum Hall phase in graphene on {SrTiO}$_3$}},\ }\href
  {https://doi.org/10.1126/science.aax8201} {\bibfield  {journal} {\bibinfo
  {journal} {Science}\ }\textbf {\bibinfo {volume} {367}},\ \bibinfo {pages}
  {781} (\bibinfo {year} {2020})}\BibitemShut {NoStop}%
\bibitem [{\citenamefont {Stepanov}\ \emph {et~al.}(2020)\citenamefont
  {Stepanov}, \citenamefont {Das}, \citenamefont {Lu}, \citenamefont
  {Fahimniya}, \citenamefont {Watanabe}, \citenamefont {Taniguchi},
  \citenamefont {Koppens}, \citenamefont {Lischner}, \citenamefont {Levitov},\
  and\ \citenamefont {Efetov}}]{Stepanov2020}%
  \BibitemOpen
  \bibfield  {author} {\bibinfo {author} {\bibfnamefont {P.}~\bibnamefont
  {Stepanov}}, \bibinfo {author} {\bibfnamefont {I.}~\bibnamefont {Das}},
  \bibinfo {author} {\bibfnamefont {X.}~\bibnamefont {Lu}}, \bibinfo {author}
  {\bibfnamefont {A.}~\bibnamefont {Fahimniya}}, \bibinfo {author}
  {\bibfnamefont {K.}~\bibnamefont {Watanabe}}, \bibinfo {author}
  {\bibfnamefont {T.}~\bibnamefont {Taniguchi}}, \bibinfo {author}
  {\bibfnamefont {F.~H.~L.}\ \bibnamefont {Koppens}}, \bibinfo {author}
  {\bibfnamefont {J.}~\bibnamefont {Lischner}}, \bibinfo {author}
  {\bibfnamefont {L.}~\bibnamefont {Levitov}},\ and\ \bibinfo {author}
  {\bibfnamefont {D.~K.}\ \bibnamefont {Efetov}},\ }\bibfield  {title}
  {\bibinfo {title} {{Untying the insulating and superconducting orders in
  magic-angle graphene}},\ }\href {https://doi.org/10.1038/s41586-020-2459-6}
  {\bibfield  {journal} {\bibinfo  {journal} {Nature}\ }\textbf {\bibinfo
  {volume} {583}},\ \bibinfo {pages} {375} (\bibinfo {year}
  {2020})}\BibitemShut {NoStop}%
\bibitem [{\citenamefont {Ciepielewski}\ \emph {et~al.}(2024)\citenamefont
  {Ciepielewski}, \citenamefont {Tworzyd\l{}o}, \citenamefont {Hyart},\ and\
  \citenamefont {Lau}}]{zenodo}%
  \BibitemOpen
  \bibfield  {author} {\bibinfo {author} {\bibfnamefont {A.~S.}\ \bibnamefont
  {Ciepielewski}}, \bibinfo {author} {\bibfnamefont {J.}~\bibnamefont
  {Tworzyd\l{}o}}, \bibinfo {author} {\bibfnamefont {T.}~\bibnamefont
  {Hyart}},\ and\ \bibinfo {author} {\bibfnamefont {A.}~\bibnamefont {Lau}},\
  }\bibfield  {title} {\bibinfo {title} {Transport effects of twist-angle
  disorder in mesoscopic twisted bilayer graphene},\ }\bibfield  {journal}
  {\bibinfo  {journal} {zenodo.10886941}\ }\href
  {https://doi.org/10.5281/zenodo.10886941} {10.5281/zenodo.10886941} (\bibinfo
  {year} {2024})\BibitemShut {NoStop}%
\end{thebibliography}%

\appendix

\renewcommand{\thefigure}{A\arabic{figure}}
\setcounter{figure}{0}    

\renewcommand{\theequation}{A\arabic{equation}}
\setcounter{equation}{0}  

\section{Generation of twist-angle domains with smooth boundaries}
\label{app:generation_samples}

We aim to create ensembles of TBG samples with randomized twist-angle domains. As a starting point, we take a rectangular-shaped pristine sample with a five-to-one proportion between its length and width. Later on, we can use a linear transformation to deform this rectangle, to make a parallelogram that conforms to the shape of the overlap region in Fig.~\ref{fig:setup}. We want to break this sample into $N_d$ domains, which are similar to  the smooth bubble-like twist-angle domains in grown TBG samples. Moreover, we wish to avoid cases in which the difference in the size of the domains is very large, something that would raise questions regarding the physical meaning of having $N_d$ distinct domains. Thus, we demand that the domains have at least an area of $10\%$ of the domain average ($A_s / N_d$, where $A_s$ is the area of the whole sample).

\begin{figure}
\includegraphics[width=0.5\textwidth]{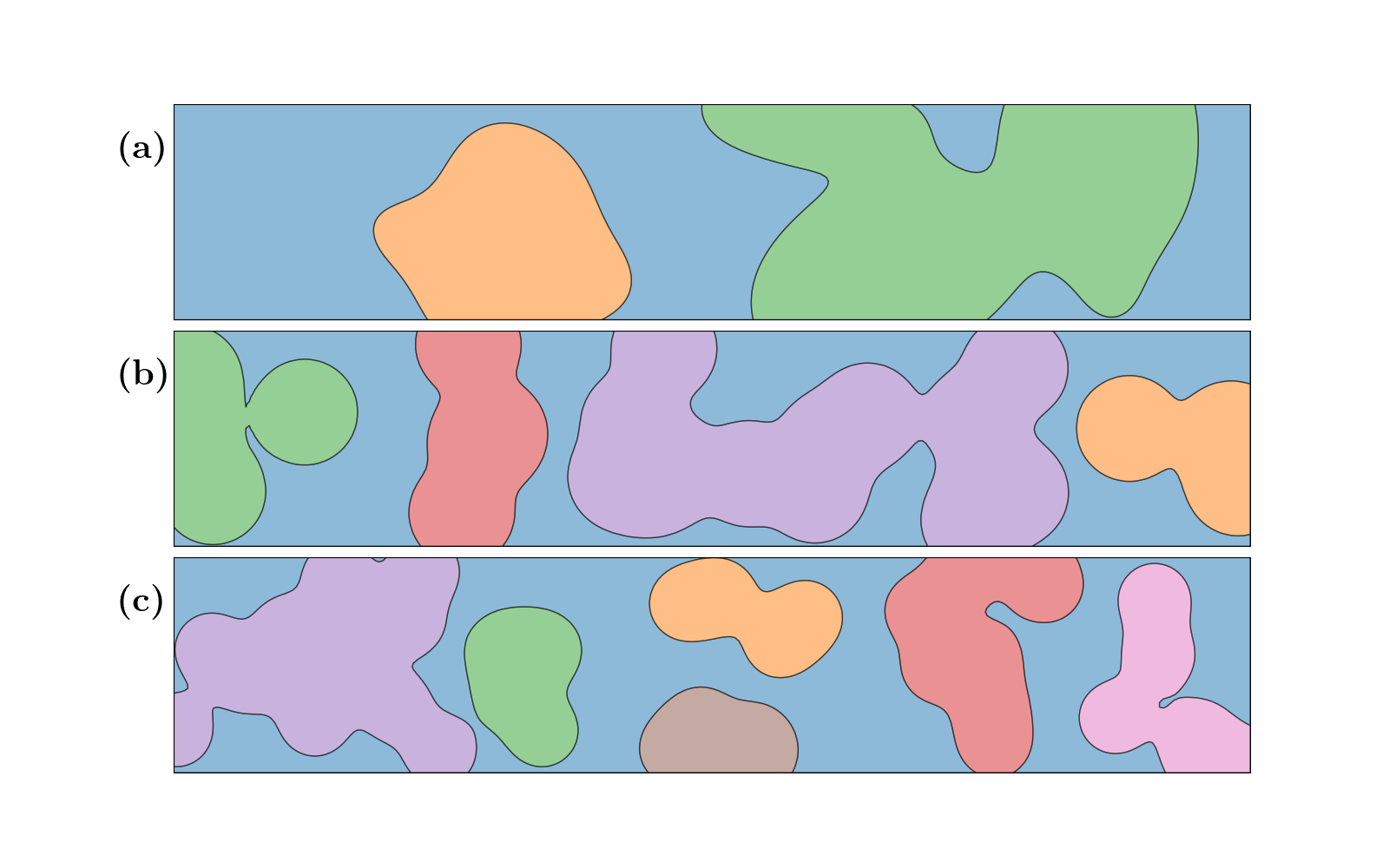}
\vspace{-1cm}
\caption{Three representative examples of rectangular pristine samples broken into $N_d$ domains with smooth boundaries: (a) $N_d = 3$, (b) $N_d = 5$ and $N_d = 7$. Each color represents one domain.}
\label{fig:domains}
\end{figure}

We find that smooth bubble-like boundaries for the domains can be created by utilizing the contour lines  $g(\vec{x}) = C$ of a scalar function 
\begin{equation}
    g(\vec{x}) = \sum_{i = 1}^{N} \frac{1}{\sigma \sqrt{2 \pi}}e^{\frac{|\vec{x} - \vec{x}_i|^2}{2\sigma^2}}, \label{contour-eq}
\end{equation}
consisting of a sum of $N$ 2D Gaussian functions, where the centers $\vec{x}_i$ are sampled from uniform random distribution and the standard deviations $\sigma$ are equal to each other. We thus define a domain as a region in the sample enclosed by such a contour line and the boundaries of the sample. Domains that are fully interior to a another domain, i.e. holes in the bigger domain, are deleted (the hole is filled and the area added to the containing domain). Finally, we define the area which remains after taking out all domains defined above from the sample as a distinct domain, even if it is composed of separated regions. Figure~\ref{fig:domains} shows three samples, with 3 (panel (a)), 5 (panel (b)) and 7 (panel (c)) domains. The residual domain described above is shown in blue.

To obtain the randomized samples we consider the ranges of parameters $N \in [20, 50]$, $\sigma \in [1.0, 5.0]$ and $C \in [0.01, 0.2]$ of Eq.~(\ref{contour-eq}). We then fix the desired number of domains to $N_d = 3, 5$ or $7$, and generate randomized samples with $N_d$ domains until we have an ensemble of 20. Samples with other $N_d$, or which contain very small domains (less than $10\%$ of the domain average) are discarded. Therefore, for each set of values $N$, $\sigma$ and $C$, we obtain an ensemble of 20 randomized samples. 

Finally, we want to have domains that are similar in size. For this purpose, for a given $N_d$, of all the ensembles generated we select the one with the smallest standard deviation $\sigma_a$ of the area of domains. The resulting three ensembles are the ones we utilized for the results shown in Fig.~\ref{fig:coductance_twist_angle_disorder}. 

We note that we discard samples having small domains because these domains would affect very little the conductance of the sample, and thus obscure the physical meaning of having an ensemble of samples with exactly $N_d$ domains. Furthermore, sharp and small domains are not typically found in real TBG samples. Furthermore, one could propose the deletion of small domains instead of repetition of the randomization process, but this would increase the variance of the area in the ensembles, as possibly multiple small domains would be added to the already typically largest residual domain.

\end{document}